\documentclass[hidelinks,12pt]{article}
\usepackage{epsfig,amsfonts,amssymb}

\usepackage{comment}
\input epsf.sty
\usepackage{cite}
\usepackage{epsfig,amssymb,euscript,xspace,xcolor}
\usepackage{amsmath,mathtools,empheq,amsthm,hyperref,graphicx}
\usepackage{mathrsfs,float}  
\usepackage{pgf,tikz,pgfplots}
\usetikzlibrary{arrows}
\usepackage[T1]{fontenc} 
\usepackage{tikz,caption,subcaption,marvosym} 
\usetikzlibrary{decorations.markings,arrows,snakes}
\usepackage[belowskip=-15pt,aboveskip=0pt]{caption}
\usepackage[skip=10pt]{caption} 
\usepackage{comment}

\usepackage[utf8]{inputenc}

\usepackage[titles]{tocloft}

\definecolor{lightblue}{rgb}{.1,.4,.5}
\definecolor{brown1}{rgb}{.64,.43,.138}

\usepackage{hyperref,cite}
\hypersetup{linktocpage, colorlinks=true,linkcolor= blue,citecolor=blue,urlcolor=lightblue}

\def\ZZZ{{\hbox{ Z\kern-1.6mm Z}}}
\def\RRR{{\hbox{ R\kern-2.4mm R}}}
\def\CCC{{\hbox{ C\kern-2.0mm C}}}
\def\zzz{{\hbox{z\kern-1mm z}}}

\newcommand{\qeq}{{\hbox{=\kern-2.3mm ? \kern.5mm }}}
\renewcommand{\qeq}{=}
\usepackage{tikz}

\newcommand{\be}{\begin{equation}}
\newcommand{\ee}{\end{equation}}
\newcommand{\ben}{\begin{eqnarray}\displaystyle}
\newcommand{\een}{\end{eqnarray}}

\def\one{{\hbox{ 1\kern-.8mm l}}}
\def\zero{{\hbox{ 0\kern-1.5mm 0}}}

\renewcommand{\theequation}{\thesection.\arabic{equation}}
\renewcommand{\theequation}{\arabic{equation}}

\newcommand{\bea}[1]{\begin{eqnarray}\label{#1} }
\newcommand{\eea}{\end{eqnarray}}

\usepackage{bm}

\def\bea{\begin{eqnarray}}
\def\eea{\end{eqnarray}}
\def\be{\begin{equation}}
\def\ee{\end{equation}}

\definecolor{wvvxds}{rgb}{0.396078431372549,0.3411764705882353,0.8235294117647058}
\definecolor{dbwrru}{rgb}{0.8588235294117647,0.3803921568627451,0.0784313725490196}
\definecolor{dtsfsf}{rgb}{0.8274509803921568,0.1843137254901961,0.1843137254901961}
\definecolor{wrwrwr}{rgb}{0.3803921568627451,0.3803921568627451,0.3803921568627451}
\definecolor{cqcqcq}{rgb}{0.7529411764705882,0.7529411764705882,0.7529411764705882}
\definecolor{rvwvcq}{rgb}{0.08235294117647059,0.396078431372549,0.7529411764705882}


\begin{document}

\baselineskip 24pt

\begin{center}

{\Large \bf  The positive geometry for $\phi^{p}$ interactions}

\end{center}

\vskip .5cm
\medskip

\vspace*{4.0ex}

\baselineskip=18pt

\centerline{\large \rm  Prashanth Raman }

\vspace*{4.0ex}

\centerline{\it ~Institute of Mathematical Sciences, Taramani, Chennai 600 113, India}
\centerline{\it ~Homi Bhabha National Institute, Anushakti Nagar, Mumbai 400085, India}

\vspace*{1.0ex}
\centerline{\it ~E-mail : prashanthr@imsc.res.in } 
\vspace*{5.0ex}
\centerline{\bf Abstract} \bigskip

Starting with the seminal work of Arkani-Hamed et al \cite{Arkani-Hamed:2017mur},  in \cite{Banerjee:2018tun}, the ``Amplituhedron program'' was extended to  analyzing (planar) amplitudes in massless $\phi^{4}$ theory. In this paper we show that the program can be further extended to include  $\phi^{p}$ ($p>4$) interactions. We show that tree-level planar amplitudes in these theories can be obtained from geometry of polytopes called accordiohedron which naturally sits inside kinematic space. As in the case of quartic interactions the accordiohedron of a given dimension is not unique, and  we show that a weighted sum of residues of the canonical form on these polytopes can be used to compute scattering amplitudes. We finally provide a prescription to compute the weights and demonstrate how it works in various examples.

\vfill \eject

\baselineskip=18pt

\tableofcontents

\newpage
\section{Introduction}

In \cite{Arkani-Hamed:2017mur, Arkani-Hamed:2013jha}  a space-time independent formulation for scattering amplitudes in ${\cal N} = 4$ SYM was proposed. A key feature of this formulation was that scattering amplitudes were to be thought of as differential forms rather than functions and it was shown that there is a precise connection between an object called the amplituhedron living in an auxiliary space and planar scattering amplitudes in ${\cal N} = 4$ SYM. It was further shown that unitarity and locality emerge quite naturally from geometric properties of the amplituhedron. Understanding scattering amplitudes in terms on canonical form on the amplituhedron makes the dual conformal symmetry manifest and has already produced striking results regarding positivity properties of the amplitudes as well as computational simplification for higher loop amplitudes \cite{Arkani-Hamed:2017tmz, Arkani-Hamed:2018rsk}.

Recently  Arkani-Hamed et al \cite{Arkani-Hamed:2017mur} extended this formulation to non-supersymmetric domain by analysing tree-level amplitudes of bi-adjoint scalar $\phi^{3}$ theory. A precise connection was established between a polytope called the associahedron and scattering amplitude of the theory. It was shown that associahedron (which is a combinatorial polytope) can be embedded as a convex polytope in the kinematic space and the canonical form associated to it is the scattering form for $\phi^{3}$ theory.  It was also shown that various properties like soft limits and recursion relations of scattering amplitudes could be obtained from geometric properties of the associahedron. This formulation was further extended to include 1-loop amplitudes in bi-adjoint $\phi^{3}$ theory \cite{Salvatori:2018fjp, Salvatori:2018aha}. 

A natural question to ask is for what class of theories such a formulation exists. In particular for theories containing independent quartic and higher order vertices, does this picture relating scattering amplitudes to differential forms and convex polytopes continues to hold.  
 First step in this direction was taken in \cite{Banerjee:2018tun} where it was shown that for tree-level planar amplitudes in $\phi^{4}$ theory such a formulation does indeed exist. It was established in \cite{Banerjee:2018tun} that there is a relationship between the scattering amplitude, scattering forms and objects called Stokes polytope. There were however important structural differences between the amplituhedron/associahedron picture and the picture that emerged in the case of quartic interactions. Namely, for an $n$-particle scattering, the corresponding geometry was a union of many polytopes of a given dimension and one had to account for all of them to capture complete tree-level planar  scattering amplitude.
 
In this paper, we extend these ideas to all $\phi^{p}$ theories by establishing a connection between \emph{unique} planar scattering forms and convex polytopes called accordiohedron. We also show that (just as in the case of cubic and quartic interactions), geometric realisation of accordiohedron inside the kinematic space leads one to identify scattering amplitudes in such theories with canonical top forms of the accordiohedron.

One of the key motivations for our work (apart from it's intrinsic interest in developing the Amplituhedron program further) was already outlined in \cite{Banerjee:2018tun} and arises from trying to  better understand the CHY formulation for $\phi^{p}$ theories \cite{Cachazo:2013hca, Baadsgaard:2015ifa}. Unlike bi-adjoint scalar field theory, or Yang-Mills theory the CHY integrands associated to $\phi^{p}$ theories are not d-log forms and hence do not seem to have a clear geometric meaning in terms of forms on the moduli space. Our hope is that by first understanding the structure of such forms in kinematic space, we may be able to better understand CHY formula for such theories. 

The rest of the paper is organised as follows. In section (2) we define the  notion Q-Compatibility and use it to  define the accordiohedron for $\phi^{p}$ interactions. In section (3) we explain how to embed the accordiohedron in kinematic space and how to obtain the scattering amplitude from the accordiohedron by intorducing {\it primitives} and {\it weights} to simplify our computations. In section(4) we derive a formula for the number of primitves of a given dimension $n$ and provide a classification all primitves up to $n\le 3$,  we also provide general prescription for determining the weights. In section (5) we prove factorization for accordiohedra and also discuss the role of factorisation in determining the weights. We finally end with some conclusions and future directions. 

\section{Amplitudes and Accordiohedron }
We shall begin by defining an object called the {\it accordiohedron} \cite{Manneville:2019, garver} associated with general dissections of polygons.  We shall then show how this reduces to the associahedron and Stokes polytopes \cite{Arkani-Hamed:2017mur, Banerjee:2018tun} for cubic and quartic interactions respectively. We shall then argue that the {\it accordiohedron} is thus the natural candidate for the positive geometry of  all $\phi^{p}$ interactions with $p\ge3$.

\subsection{Accordion lattices and Accordiohedra} \label{accord}
\begin{figure}[ht!] 
\centering 
\includegraphics [scale=0.4] {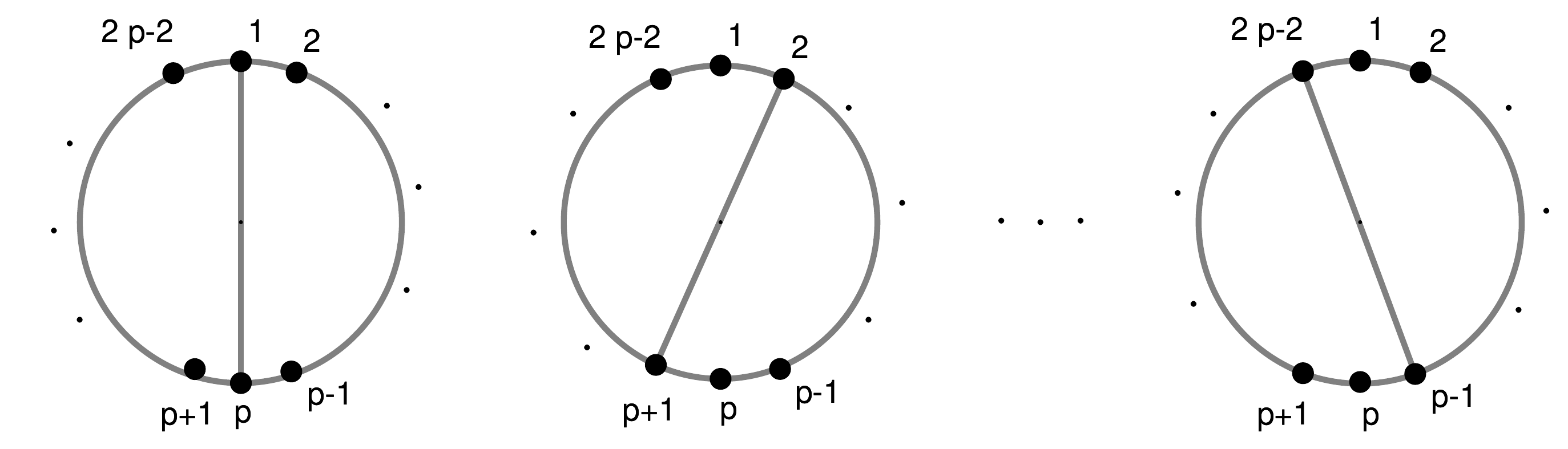} 
\setlength{\belowcaptionskip}{8pt}
\caption{The p-1 $p$-angulations of A}
\label{fig: polyminoes} 
\end{figure}

Let A be a convex polygon. Let us consider the division of A into identical $p$-gons which we call $p$-angulation of A. We can represent A as a set of points on the unit circle oriented clockwise where the arcs represent edges of A and chords represent diagonals of A. The simplest example is the case where we divide $(2p-2)$-gon A into two $p$-gons (see figure (\ref{fig: polyminoes})). There are  $p-1$ possible $p$-angulations which correspond to having the diagonals $ \{(1,p),(2, p+1), \cdots ,(p-1, 2p-2) \}$. 
We define a notion of {\it Q-compatible diagonal} as \footnote{In \cite{Manneville:2019, garver} there is different definition of compatability, but these two definitions can be shown to be equivalent to each other and we shall use the definition (\ref{Qcompat}) as its most suited for our purposes. We thank Alok Laddha for explaining this fact to us. }:
   \bea \label{Qcompat}
 (i ,j) \rightarrow (Mod(i+p-2, 2p-2),Mod(j+ p-2, 2p-2)) 
 \eea 
 We can use this rule to define  accordion lattices ${\mathcal AL_{p,n}^{P}}$ of dimension $n$ associated with  a reference   ~$p$ -angulation P \footnote{We consider only the case where we divide the polygon into $p$-gons in this paper, but accordion lattices are defined for arbitrary dissections.}  as follows: 
 
 \begin{itemize}
\item {\it We can start with any $p$-angulation $P$ of a convex polygon with $n$ diagonals.}

\item {\it In the first step for each of the $n$ diagonals, we go to the unique $(2p-2)$-gon which contains it and replace it with its $Q$-compatible diagonal.}

\item {\it In the second step for each of the $n$ $p$-angulations at the end of step one we choose one of the original $n-1$ diagonals and replace it with its $Q$-compatible diagonal as in step one.}

\item {\it We repeat this till none of the original $n$ diagonals remain in step $n$. } 
 
 \end{itemize}
This generates a graph\footnote{More precisely one can define a partial order on the the set of all $p$-angulations that can be obtained from a particular $p$-angulation P, which has a maximal and a minimal element ($P$ itself) thus making it lattice and this procedure generates the Hasse diagram of the lattice } which is the 1-skeleton of a convex polytope called the {\it Accordiohedron} \cite{Manneville:2019, garver}, which we shall also  call ${\mathcal AC_{p,n}^{P}}$.\footnote{Interestingly enough, there is yet another, ``complimentary'' family of polytopes known as graph associahedra \cite{DEVADOSS2009271} which also had deep connections with geometry of scattering amplitudes. In contrast to accordiohedron, graph associahedra can not always be obtained by considering dissections of polygons. graph associahedra is a set of polytopes which includes, associahedron, permutahedron, halohedron etc. Many of these members, e.g. permutahedron and halohedron are associated to amplitudes in bi-adjoint scalar theory with non-planar \cite{He:2018pue} and 1-loop amplitudes \cite{Salvatori:2018aha} respectively.  It is intriguing that one class of polytopes helps one to move beyond tree-level and planarity in bi-adjoint $\phi^{3}$ theories and the other class helps one move beyond cubic vertices.}

 \bea
  {\bf Vertices} &\leftrightarrow& {\bf Q \text{-}compatible ~p\text{-}angulations } \nonumber \\
  {\bf Edges} &\leftrightarrow & {\bf Flips ~between ~ them } \nonumber \\
  {\bf k\text{-}Facets} &\leftrightarrow& {\bf k\text{-}partial ~p\text{-}angulations} \nonumber 
  \eea

\begin{figure}[h]
\centering 
\includegraphics [scale=0.4] {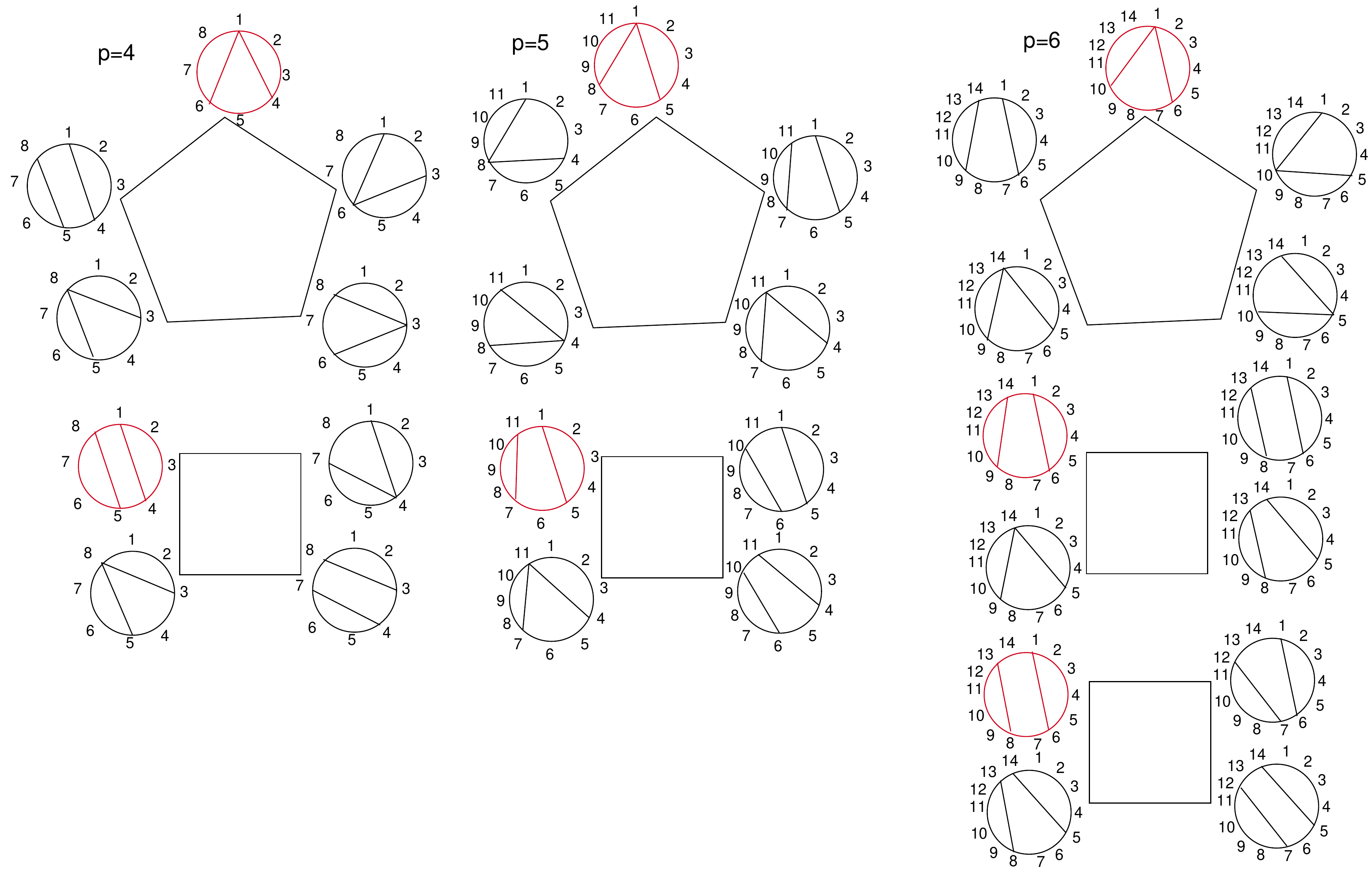} 
\setlength{\belowcaptionskip}{8pt}
\caption{accordiohedra for the n=2 case. The red circles indicate the reference $p$-angulations. }
\label{fig: accordion2} 
\end{figure}
In the case of cubic interactions ($p=3$), (\ref{Qcompat}) reduces to $(i,j) \rightarrow (Mod(i+1,4),Mod(j+1,4))$ which is the usual {\it mutation} rule and the resulting accordiohedron ${\mathcal AC_{3,n}^{P}}$ is the associahedron \cite{Arkani-Hamed:2017mur}.
   
In the case of quartic interactions ($p=4$), (\ref{Qcompat}) reduces to  $(i, j) \rightarrow (Mod(i+2, 6), Mod(j+ 2, 6 ))$ which was the {\it Q compatibility} rule defined in \cite{Baryshnikov}  and the accordiohedron ${\mathcal AC_{4,n}^{P}}$  was shown to be the Stokes polytope \cite{Banerjee:2018tun}.
  
Thus the Accordiohedra ${\mathcal AC_{p,n}^{P} }$ are a general class of polytopes which contain both associahedron and Stokes polytopes as special cases when the $p$-angulations corresponds to triangulations and quadrangulations respectively. The accordiohedra ${\mathcal AC_{p,n}^{P} }$ with $p>4$ also retains many of the features of the Stokes polytopes we had discussed earlier in \cite{Banerjee:2018tun} including the fact that the accordiohedron  ${\mathcal AC_{p,n}^{P} }$  of a given dimension $n$  is not unique and depends on the reference $p$-angulation P, which is due to the fact that (\ref{Qcompat}) is not an equivalence relation as $(1,p) \rightarrow (p-1,2p-2)$, but $(p-1,2p-2) \rightarrow (p-2,2p-3) \neq (1,p) $  except when $p=3$. 

The case $p=3$ is special in this sense as for ${\mathcal AC_{3,n}^{P} }$ is independent of $P$, as every diagonal is $Q$-compatible with every other diagonal and thus we could start with any triangulation $P$ and we would generate all possible triangulations.

 The accordiohedron obtained by starting with a particular $p$-angulation is also completely determined by the relative configuration of diagonals.\footnote{From the perspective of Feynman graphs this is equivalent to saying that there is an accordiohedron for each topological class of graphs.}
 
The $n=1$ accordiohedron  ${\mathcal AC_{p,1}^{(i,p+i)}}$ are  lines with vertices $(i,p+i)$ and $(Mod(i+p-2,2p-2),Mod(i+2p-2, 2p-2))$ for $i=1,...,p-1$.

 The $n=2$ case the accordiohedron can be either pentagons or squares  depending on whether the two diagonals meet or don't meet respectively (see fig(\ref{fig: accordion2})) just as in the case of Stokes polytopes. In other words ${\mathcal AC_{p,2}^{(P)}} \equiv {\mathcal AC_{4,2}^{(Q)}} $ for all $p$ provided both $P$ and $Q$ have the same configuration of diagonals, we shall prove this in a later section (\ref{2case}) by establishing the precise maps between vertices of the Stokes polytope and that of the accordiohedron. 
 
 The $n=3$ case the accordiohedra continue to be one of the four  $n=3$ Stokes polytopes with different multiplicities, {\it i.e.}, ${\mathcal AC_{p,3}^{(P)}} \equiv {\mathcal AC_{4,3}^{(Q)}} $ for all $p$ provided $P$ and $Q$ have the same configuration of diagonals. We elaborate on this in section (\ref{3case}). We expect that at higher $n$ new polytopes which are not one of the Stokes polytopes will be eventually generated.

\section{Positive geometry for $\phi^{p}$ interactions}
We would like to show that the accordiohedron  ${\mathcal AC_{p,n}^{(P)}}$  is the positive geometry associated to $\phi^{p}$ interactions.  We shall do this by first embedding the accordiohedron into kinematic space and then showing that the canonical form of the accordiohedron when pulled back  gives the right planar scattering amplitude for $\phi^{p}$ interactions.
We start by noting the following facts:
\begin{itemize}
\item The only tree level amplitudes consistent with $\phi^p$ interactions have $p+(p-2)n$ external legs for $n+1$ vertices. 
\item Analogously to the cubic and quartic cases there is a 1-1 correspondence between planar tree level Feynman graphs and dissections of $p+(p-2)n$- gon into p-gons.
\item We also require the accordiohedron ${\mathcal AC_{p,n}^{(P)}}$ to have dimension $n$, which is the number of propagators.\footnote{This is because we require the top-form on the positive geometry, once embedded in kinematic space to produce the right scattering amplitude.}
\end{itemize}
We shall first introduce some notations and variables that we shall use to describe the kinematic space.
\subsection{Kinematic space}
We shall primarily follow the conventions of \cite{Arkani-Hamed:2017mur} which we briefly summarise here.

The kinematic space ${\mathcal K_n}$ of $n$ massless momenta in $d$ dimensions (with $n< d-1$) is spanned by the Mandelstram variables:
\bea
s_{ij} &=&(p_i+p_j)^{2}, ~~~ ~~~~~ {\rm with} ~1\le i< j\le n \nonumber
\eea
that satisfy momentum conservation
\bea
\sum_{{j=1}\atop{j \ne i}}^{n} s_{ij} =0 \nonumber
\eea
For any set of particle labels $I \subset \{ 1,2,...,n\}$ we can define generalised Mandelstram variables as:
\bea
 s_I &=& \left( \sum_{i \in I} p_i \right)^{2} = \sum_{{i,j \in I}\atop{i<j} } s_{ij} \nonumber
\eea
A more convenient basis for the kinematic ${\mathcal K_n}$ for our purposes are the planar kinematic variables,
\bea
X_{ij}& =&s_{i,i+1,...,j-1}, ~~~~~~~~{\rm with} ~~~ 1\le i< j\le n \nonumber
\eea
It is clear that $X_{i~i+1}=0$ and $X_{1n} =0$ due to masslessness and momentum conservation respectively.
The mandelstram variables $s_{ij}$ can be written interns of planar variables $X_{ij}$ as
\bea \label{mantoplan}
s_{ij} = X_{i~ j+1}+X_{i+1 ~j}-X_{i ~j}-X_{i+1~ j+1}
\eea
The planar variables $X_{ij}$ can be interpreted as the the length of a diagonal between vertices $i$ and $j$ of an $n$-gon which has the momenta $p_i$ as sides.

\subsection{Planar scattering form for $\phi^{p}$ interactions}
We would like to define a planar scattering form for  $\phi^p$ interactions. We can associate to each planar graph $g$ with propagators $X_{i_1 j_1}, X_{i_2 j_2} \cdots X_{i_n j_n} $ a scattering form:
\bea
 \frac{\sigma(g)}{\prod_{k=1}^{n}  X_{i_k j_k}}  d X_{i_1 j_1} \wedge d X_{i_2 j_2} \wedge \cdots \wedge  d X_{i_n j_n} \nonumber
\eea
where, $\sigma(g) = \pm 1$. 

Thus, when we sum over all planar graphs $g$ we have several possible scattering forms. Since we do not have a notion of {\it projectivity } except in the case of $p=3$ which helps us fix a unique scattering form \cite{Arkani-Hamed:2017mur}. We can choose a particular  reference graph $g$ (equivalently a $p$-angulation $P$) and look at only those subset of graphs which are related to this graph by a sequence of $Q$-flips namely all the vertices of the {\it accordiohedron}. If a graph $g'$ is related to $g$ by an odd (even) number of $Q$-flips we can associate $-(+)$ sign to it. Thus, we can define a $p$-angulation $P$ dependent planar scattering form $\Omega_{n}^P$ :
\bea
\Omega_{n}^P = \sum_{flips} \frac{(-1)^{\sigma(flip)}} {\prod_{k=1}^{n}  X_{i_k j_k}}  d X_{i_1 j_1} \wedge d X_{i_2 j_2} \wedge \cdots \wedge  d X_{i_n j_n}  \nonumber
\eea
Since, the $Q$-compatible $p$-angulations corresponding to any reference $p$-angulation $P$ does not exhaust all the $p$-angulations, we need to define such a planar scattering form for each $P$. 

In the $n=1$ case the set of $Q$-compatible p-angulations are $ \{ (1~p;+),(p-1~2p-2 ;-) \}, \{ (2~p+1;+),(p~2p-1;-) \}, \cdots, \{ (p-1~2p-2;+),(p-2~2p-3;-) \}$\footnote{Here the signs denote $(-1)^{\sigma(flip)}$, when we have multiple diagonals we need to carefully maintain the order of diagonals when we flip as it contibutes to the sign.} the planar scattering forms for which are:
\bea
\Omega_ {2p-2}^{(ij)} =d \ln X_{i~j} - d\ln X_{i+p-2 ~ j+p-2} \nonumber 
\eea
where, $i,j =1,...,p-1$ modulo $(2p-2) $ with $|i-j|=p-1$.

We now turn to embedding the accordiohedron in kinematic space and showing that when the planar scattering form is pulled back onto the accordiohedron it gives the canonical form of the accordiohedron.

\subsection{Locating the accordiohedron  inside kinematic space}
We now define the kinematic accordiohedron ${\mathcal AC_{p,n}^{(P)}}$. We locate the {\it accordiohedron} inside the positive region of kinematic space $X_{ij} >0$ for all $1\le i<j \le p+(p-2)n$ by imposing the following constraints:
 \bea \label{embed}
s_{ij} &=& - \, c_{ij} \ ;  \quad  for \ \ 1\le i < j \le p-1+(p-2)n , \ |i-j| \geq 2    \nonumber \\
X_{r_i s_i} &=& d_{r_i s_i}  \ ;  ~ s.t.  ~ {P} \cup_{i=1} ^{n} \{(r_i, s_i) \}~ {\rm is ~a ~complete ~triangulation} 
\eea
where $c_{ij},d_{r_i s_i}$ are positive constants.\footnote{In the case when $p=4$ that is when the Accordiohedron is a Stokes polytope, there is a canonical choice for the additional constraints if the Stokes polytope is itself not an Associahedron \cite{Banerjee:2018tun}. 
However for $p\ >\ 4$ we do not have any canonical choice of these constraints.  As we show in this section, there is at least one choice which consistently embeds the Accordiohedron in the Kinematic space.} 

Physically we choose the above set of constraints as they do not appear as propagators of any $\phi^{p}$ graph. The first constraint above is the famous associahedron embedding \cite{Arkani-Hamed:2017mur}. We have thus embedded the accordiohedron inside the associahedron. The positivity of $X_{ij}$'s, the above constraints along with the equation (\ref{mantoplan}) are a set of inequalities satisfied by the $X_{ij}$  which makes the convexity of the accodiohedron manifest. 

We first consider the $n=1$ case with the reference $p$-angulation to be $P= \{ (1, p) \} $ for $p=5,6$. For $p=5,6$ we can choose $  \cup_{i} \{r_i~s_i \}$ to be  $\{ 24,25,17,57  \}$ and  $\{ 35,36,26,17,19,79 \}$  respectively. The above constraints then translate to: 

{\bf p=5}: ~~~ $X_{48} =  \sum_{i=1}^{3}c_{i5}+c_{i6}+c_{i7} -X_{15}  $ which a line with boundaries at $X_{15},X_{48} =0$ 

provided the following are satisfied \footnote{Since we are slicing the associahedron using some hyperplanes $X_{r_i s_i} = d_{r_i s_i} $ to get the accordiohedron these constraints tell us how the slicing should be made. For higher $n$ we shall not state these constraints for brevity but we shall assume that they are satisfied.}
\begin{center}
 $~~~~~~~~~\sum_{i=1}^{3} c_{i7} \le  d_{17} \le \sum_{i=1}^{5} c_{i7}$  \\
$~~~~~~~~~~~~~~~~~~~~~~~~\sum_{i=5}^{7}( c_{2i}+c_{3i})  \le  d_{25} \le \sum_{i=3}^{7} c_{1i}+ \sum_{i=5}^{7}( c_{2i}+c_{3i})$  \\
$~~~~~~~~~~~~~~~~~~~0 \le d_{24} \le d_{25} +c_{24}$ \\
$~~~~~~~~~~~~0 \le d_{57} \le c_{46} $ \\
\end{center}

{\bf p=6}: ~~~ $X_{510} =  \sum_{i=1}^{4}c_{i6}+c_{i7}+c_{i8}+c_{i9} -X_{16}  $ which a line with boundaries at $X_{16},X_{510} =0$ 

provided the following are satisfied
\bea
 \sum_{i=1}^{4} c_{i9} &\le&  d_{19} \le \sum_{i=1}^{7} c_{i9}, ~~~~~~~~~~~~~~~~~~~~~~~
 \sum_{i=1}^{4} \sum_{j=7}^{9} c_{ij} \le  d_{17} \le \sum_{i=1}^{4} \sum_{j=7}^{9} c_{ij} +d_{19}  \nonumber\\
 \sum_{i=2}^{4} \sum_{j=6}^{9} c_{ij} &\le&  d_{26} \le \sum_{i=2}^{4} \sum_{j=6}^{9} c_{ij} + \sum_{i=3}^{9} c_{1i}, ~~~~~~
 \sum_{i=3}^{4} \sum_{j=6}^{9} c_{ij} \le  d_{36} \le c_{24}+c_{25}+d_{26}  \nonumber\\
0 &\le&   d_{79} \le \sum_{i=1}^{6} c_{i8} +d_{19},  ~~~ ~~~~~~~~~~~~~~~~~~~~~~~
0 \le d_{35} \le c_{35} +d_{36}  \nonumber
\eea
The above equations define lines with $Q$-compatible vertices  $ \{15,48 \} $ and $ \{ 16,510\} $ for $p=5$ and $p=6$ respectively. We can trivially repeat this exercise for any other reference p-angulation $P =\{ i,i+p\}$, the results of which can be obtained by taking $k \rightarrow k+i-1$ in the above equations.
We can now pull back the scattering form onto the {\it accordiohedron} ${\mathcal AC_{p,n}^{P}}$ as:
\bea
\omega_{p,n}^{P} =\left(\frac{1}{X_{i ~i+p-1}} +\frac{1}{X_{i+p-2~ i+2p-3}}\right) d \ln X_{i i+p-1} :=  m_{p,n}^{P} ({ \mathcal AC_n })  d \ln X_{i i+p-1} \nonumber
\eea
with $i=1,...,p-1$.

As before to get the full amplitude $ {\cal M}_n $  we consider  a weighted sum $ \tilde{\cal{M}}_n $ of $m_{p,n}^{P}$ over all $P$.
\bea
\tilde{\cal{M} }_n = \sum_{i=1}^{p-1} \alpha_{i} \left(\frac{1}{X_{i ~i+p-1}} +\frac{1}{X_{i+p-2~ i+2p-3}}\right) 
\eea
It is clear that $ {\cal M}_n = \tilde{\cal{M}}_n $ if and only if $\alpha_{i} =\frac{1}{2}$ for all $i=0, \cdots ,p-1$.

{\it Thus, we can simplify our computation by considering a subset of $p$-angulations  $\{P_{1},\dots, P_{I}\}$ called primitive $p$-angulations for which: 

(a) no two members of the set are related to each other by cyclic permutations and

(b) all the other $p$-angulations can be obtained by a (sequence of) cyclic permutations of one of the $P$s belonging to the set.}

The primitives are the class of rotationally inequivalent diagrams. Since, a rotation does not change the relative configuration of diagonals it is clear that accodiohedra remain the same for all the diagrams that belong to a primitive class and that the weights depend only on primitives. We shall say more about primitives in section (\ref{primeticountsection}). 
\bea \label{masterformula}
{\cal{M}}_n  = \sum_{{rotations}\atop{ \sigma}} \sum_{{primitives}\atop{ P}} \alpha_{P} ~ m_{p,n}^{(\sigma.P)}
\eea
For now let us look at a couple of examples to see how finding primitive accordiohedra and their weights help us in getting the scattering amplitude.

In the $n=1$ case above there was only one primitive $ P=\{(1,p) \}$.

We consider the $n=2$ case for $p=5,6$ for which we now have the set of primitives as (see figure(\ref{fig: accordion2})) $\{ (15,610) ,(15,18)\}$ and
 $\{ (16,914) ,(16,110),(16,813)\}$ respectively.
The set of $Q$-compatible $p$-angulations for these are: 
\bea
 {\bf  p=5} : ~~S_5^{1} &=& \{ (15,711;+),(411,711;-),( 15,610;-),(411,610;+) \} ,\nonumber \\    S_5^{2} &=& \{ (15,18;+),(18,48;-),( 15,711;-),(411,711;+) ,(411,48;+)\} \nonumber \\\nonumber\\
  {\bf  p=6} : ~~S_6^{1} &=& \{ (16,914;+),(514,914;-),( 16,813;-),(514,813;+) \} ,\nonumber\\   S_6^{2} &=& \{ (16,813;+),(514,813;-),(16,712;-)( 514,712;+) \} ,\nonumber \\ S_6^{3} &=& \{ (16,110;+),(16,914;-),( 110,510;-),(510,514;+),(514,914;+) \} \nonumber  \eea \\
  The embedding constraints (\ref{embed}) can be solved to obtain: 
  
  {\bf p=5}:  For P=(15,711), with $\cup_{i} \{r_i~s_i \}$ to be $\{13,35,17,57,810,710  \}$ 
  \bea
 X_{411} &=& \sum_{i=1}^{3} \sum_{j=5}^{10} c_{ij}- X_{15} \nonumber\\
  X_{610} &=& \sum_{i=1}^{5} \sum_{j=7}^{10} c_{ij}+c_{610}-d_{17} + d_{710}- X_{711} \nonumber
  \eea
  The positivity of $X_{610},X_{411}$ carves out a square region in the $X_{15}, X_{711}$ space. The accordiohedron in this case is also a square as we had emphasised in section (\ref{accord}). \\
  For P=(15,18), with $\cup_{i} \{r_i~s_i \}$ to be $\{13,35,16,68,810,110  \}$ 
  \bea
 X_{48} &=& \sum_{i=1}^{3} \sum_{j=5}^{7} c_{ij}- X_{15} +X_{18}\nonumber\\
  X_{411} &=& \sum_{i=1}^{3} \sum_{j=5}^{10} c_{ij}- X_{15} \nonumber\\
  X_{711} &=& \sum_{i=1}^{6} \sum_{j=8}^{10} c_{ij}- X_{18} \nonumber
  \eea

  {\bf p=6}:  For P=(16,914), with $\cup_{i} \{r_i~s_i \}$ to be $\{13,35,15,911,912,913,814,17,714 \}$ 
  \bea
 X_{514} &=& \sum_{i=1}^{4} \sum_{j=6}^{13} c_{ij}- X_{16} \nonumber\\
  X_{610} &=& c_{813}+d_{814} + d_{913}- X_{914} \nonumber
  \eea
  For P=(16,813), with $\cup_{i} \{r_i~s_i \}$ to be $\{13,35,15,713,912,911,812,17,714 \}$ 
  \bea
 X_{514} &=& \sum_{i=1}^{4} \sum_{j=6}^{13} c_{ij}- X_{16} \nonumber\\
  X_{712} &=& c_{712}+d_{713} + d_{812}- X_{813} \nonumber
  \eea
  For P=(16,110), with $\cup_{i} \{r_i~s_i \}$ to be $\{13,35,15,17,18,19,113,1113,111 \}$ 
  \bea
 X_{510} &=& \sum_{i=1}^{4} \sum_{j=6}^{9} c_{ij}- X_{16} +X_{110}\nonumber\\
  X_{514} &=& \sum_{i=1}^{4} \sum_{j=6}^{13} c_{ij}- X_{16} \nonumber\\
  X_{914} &=& \sum_{i=1}^{8} \sum_{j=10}^{13} c_{ij}- X_{110} \nonumber
  \eea
  When pulled back onto constraints the corresponding $ m_{p,n}^{P} $'s are:
  
  ${\bf p=5}$:
  \bea
   m_{5,2}^{1} &=& \left( \frac{1}{X_{15}X_{ 610}}+\frac{1}{X_{411}~ X_{610}} +\frac{1}{X_{15}~X_{ 711}}+\frac{1}{X_{411}~ X_{711}}\right) d  X_{15} \wedge X_{610} \nonumber  \\
      m_{5,2}^{2} &=& \left( \frac{1}{X_{15}X_{ 18}}+\frac{1}{X_{18}~ X_{48}} +\frac{1}{X_{15}~X_{ 711}}+\frac{1}{X_{411}~ X_{711}} + \frac{1}{X_{48}~ X_{411}}\right) d  X_{15} \wedge X_{18} \nonumber
  \eea
  Plugging the above forms  into eq. (\ref{masterformula}) with weights $\alpha_{5,2}^{P_1} = \frac{3}{11}$,  $ \alpha_{5,2}^{P_2}= \frac{2}{11} $ (see section (\ref{weights}) for details) gives the right amplitude.
  
    ${\bf p=6}$:
  \bea
   m_{6,2}^{1} &=& \left( \frac{1}{X_{16}X_{ 914}}+\frac{1}{X_{514}~ X_{914}} +\frac{1}{X_{16}~X_{ 813}}+\frac{1}{X_{514}~ X_{813}}\right) d  X_{16} \wedge X_{914} \nonumber  \\
   m_{6,2}^{2} &=& \left( \frac{1}{X_{16}X_{ 813}}+\frac{1}{X_{514}~ X_{813}} +\frac{1}{X_{16}~X_{ 712}}+\frac{1}{X_{514}~ X_{712}}\right) d  X_{16} \wedge X_{813} \nonumber  \\
      m_{6,2}^{3} &=& \left( \frac{1}{X_{16}X_{ 110}}+\frac{1}{X_{16}~ X_{914}} +\frac{1}{X_{110}~X_{ 510}}+\frac{1}{X_{510}~ X_{514}} + \frac{1}{X_{514}~ X_{914}}\right) d  X_{16} \wedge X_{110} \nonumber
  \eea
 Plugging the above forms  into equation (\ref{masterformula}) with weights $\alpha_{6,2}^{P_1} = \frac{1}{3}$,  $ \alpha_{6,2}^{P_2}= \frac{1}{6} $, $\alpha_{6,2}^{P_3} = \frac{1}{3}$ (see section (\ref{weights}) for details) gives the right amplitude.
 
\section{Analysing the combinatorics of Accordiohedra} \label{primeticountsection}
Both in \cite{Banerjee:2018tun} and in the present case, a complete computation of the amplitude from the geometry of the polytope requires determination of all the primitives of a given dimension $n$ and computation of the corresponding weights. We shall address the problem in this section. We emphasise that this a purely combinatorial problem and hence does not depend on the construction of kinematic space accordiohedron. In sections (\ref{primicountquart}) ,(\ref{primitivecountp}) we shall first derive  formulae to count the number of primitive accordiohedra of a given dimension $n$. Then in sections (\ref{2case}) ,(\ref{3case}) we provide a complete classification of primitive accordiohedra  for $n\le 3$ and compute the corresponding weights for any $\phi^{p}$ interactions. Let us first consider the quartic case.

\subsection{Counting primitives for the quartic case} \label{primicountquart}
 
 In this section we shall address the quartic case first and provide a formula for the number of primitive Stokes polytopes $p_n$ of a given dimension $n$. The main result of this section is:
 \bea
p_n = \begin{cases}
\frac{1}{2n+4} F_{n+1} + \frac{1}{2} F_{\frac{n+1}{2}}, ~n = 2 k+1 \\
\frac{1}{2n+4} F_{n+1} + \frac{1}{4} \tilde{F}_{\frac{n}{2}}, ~n = 2 k~ {\rm with}~ k ~{\rm odd}  \\   \nonumber
\frac{1}{2n+4} F_{n+1} + \frac{1}{2} F_{\frac{n}{4}},~ n = 4 k
\end{cases}  
\eea
We shall now prove this result.

We shall consider a $(4+2n)$-gon as equally spaced points $a_i$ on the circle i.e. $a_i = \exp{\tfrac{2\pi i}{4+2n}}$ with $i=0,...,2n+3$. The edges of the polygon correspond to arcs $a_i a_{i+1}$ on the circle and the diagonals of a quadrangulation correspond to chords. 

There is a natural action of the dihedral group $\mathbb{D}_{2n+4}$ on any given quadrangulation which is generated by rotation and reflections about a given diagonal. We are interested in counting {\it primitive quadrangulations}, no two of which are related to each other by a cyclic permutation, which corresponds to a rotation on the circle. Thus, it is sufficient to consider only the cyclic group  $\mathbb{Z}_{2n+4}$  for our purposes. The problem of counting {\it primitive quadrangulations} is thus equivalent to finding the number orbits of the set of all quadrangulations of a $(4+2n)$-gon under the action of the cyclic group  $\mathbb{Z}_{2n+4}$. 

We shall do this by using the celebrated  {\it Burnside's lemma}, which is the  standard way to count the number of orbits $G/G_X$ for the action of any finite group $G$ on a set $X$. It states that the number of orbits is equal to the average number of points that  remain invariant when acted on by elements of $G$.
\bea
|G/G_X | = \frac{1}{|G|} \sum_{g ~\in ~G} | \{e ~\in ~X |~g.e =~e  \} | 
\eea

Thus, to count the number of {\it primitive quadrangulations} we just need to find the subset of quadrangulations that are invariant under some rotation. This problem has been addressed by \cite {HARARY} using the method of generating functions, but we shall take a simpler approach here following \cite{Nikos}.

We can consider the division of $(2n+4)$-gon into $n+1$ quadrilaterals.  We first note that the centre of the circle is left invariant by the action of the cyclic group  $\mathbb{Z}_{2n+4}$. The centre of the circle can lie on:

(1) A diameter. This can only happen when $n$ is odd since, the relative angle between the end points $a_i, a_j$ of this diameter has to $\pi$.

(2) The midpoint of an invariant cell i.e. on the point of intersection of the diagonals of a centre square which remains invariant.

\begin{figure}[h!] 
\centering 
\includegraphics [scale=0.5] {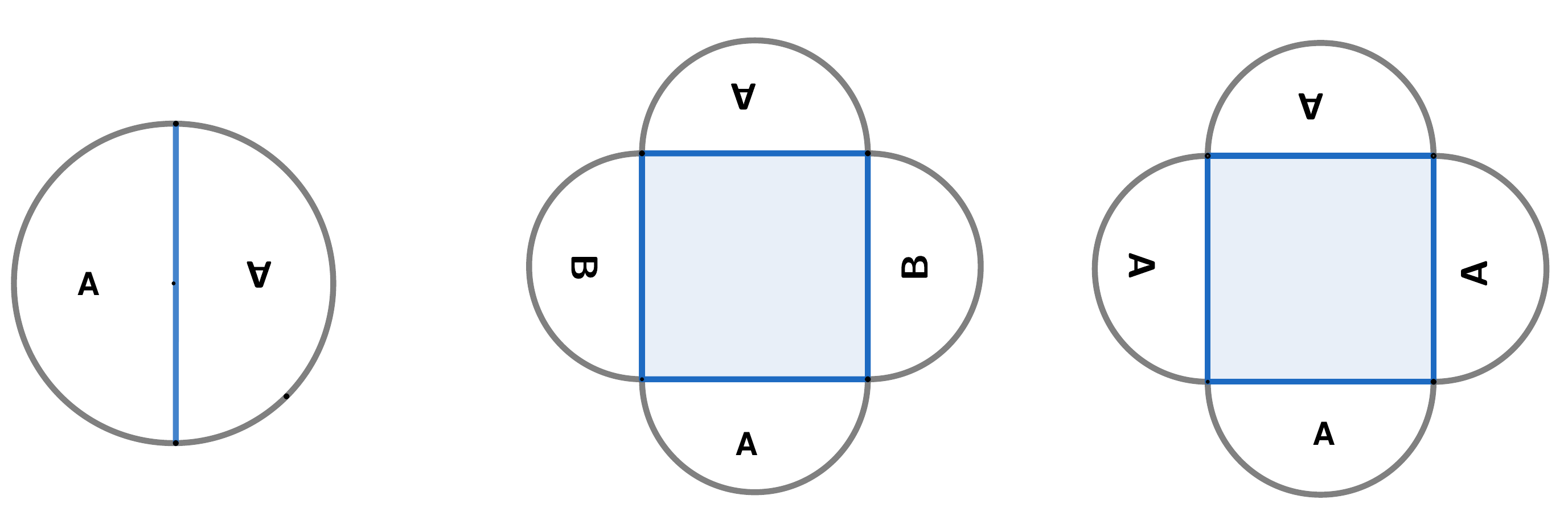} 
\setlength{\belowcaptionskip}{8pt}
\caption{All the quadrangulations invariant under some rotation}
\label{fig: prims} 
\end{figure}

 In case (1) the diameter forms an axis of symmetry and has to be left invariant by rotations and it is clear that the only possible rotation which does this is by $\pi$ and the quadrangulation $Q$ consists of a left and a right part where the left part is a rotation of the right one.( see figure(\ref{fig: prims}) ). 
 
 In case (2) the diagonals can either be rotated to themselves or into each other. This can only be accomplished by rotations of $\pi, \pm \frac{\pi}{2}$ and the corresponding  quadrangulations are shown in the figure (\ref{fig: prims}).
 
The number of quadrangualtions of $(2n+4)$-gon into $n+1$ quadrangles is given by the Fuss-Catalan number $ F_{n}=\frac{1}{2n+1} \left({3n} \atop{n}\right)$ (which we derive in appendix \ref{appendixA}). The number of quadrangulations of type (1) is $n F_{\frac{n+1}{2}}$, as we can choose a diameter in $n+2$ ways and for each choice of the diameter there $ F_{\frac{n+1}{2}}$ sub-quadrangulations $A$.

The number of quadrantulations of type (2) depends on whether $n$ is divisible by 2 or 4 and is given by $ \frac{(n+2)}{2} \tilde{F}_{\frac{n}{2}}$  and $ (n+2)F_{\frac{n}{4}}$ respectively. In the case where $n=2k$ we can divide $k$ into $k_1, k_2$  which we call $A$ and $B$ in the third figure  of (\ref{fig: prims}) and the number of such quadrangulations would correspond to $F_{k_1} $ and $F_{k-k_1} $ respectively.  The total number of such quadrangulations would then be $\sum_{k_1=0}^{k} F_{k_1} F_{k-k_1}$and since there are $\frac{n+2}{2}$ ways we can relabel the invariant square.
Using the following combinatorial identity (see appendix \ref{B}).
\bea
\tilde{F}_{k} = \sum_{k_1=0}^{k} F_{k_1} F_{k-k_1} = \frac{2}{2n+2} \left({3n+1} \atop{n}\right) \nonumber
 \eea 
We have  $ \frac{(n+2)}{2} \tilde{F}_{\frac{n}{2}}$ invariant quadrangulations under a rotation by $\pi$.

When $n=4k$ we have $F_{\frac{n}{4}}$ subquadrangulations $A$ as shown in figure (\ref{fig: prims}). There are also $(n+2)$ ways to relabel the invariant cell and thus there are a total of $ (n+2)F_{\frac{n}{4}}$ quadrangulations that are invariant under a rotation by $\pm \frac{\pi}{2}$. Thus, after also including the identity rotation which leaves all the elements invariant we get the total number of {\it primitive quadrangulations} $p_n$ is given by:

\bea
p_n = \begin{cases}
\frac{1}{2n+4} F_{n+1} + \frac{1}{2} F_{\frac{n+1}{2}}, ~n = 2 k+1 \\
\frac{1}{2n+4} F_{n+1} + \frac{1}{4} \tilde{F}_{\frac{n}{2}}, ~n = 2 k~ {\rm with}~ k ~ {\rm odd}  \\   \nonumber
\frac{1}{2n+4} F_{n+1} + \frac{1}{2} F_{\frac{n}{4}},~ n = 4 k
\end{cases}  
\eea
We can easily check the above formula for $n=1,2,3 $ cases by using $F_{n} = 1,3,12,55$ and $\tilde{ F}_{n} =1,2,7,30$ for $n=1,2,3,4$. The set of invariant quadrangulations is shown in the figure (\ref{fig: primex}) below.
  \begin{figure}[h] 
\centering 
\includegraphics [scale=0.4] {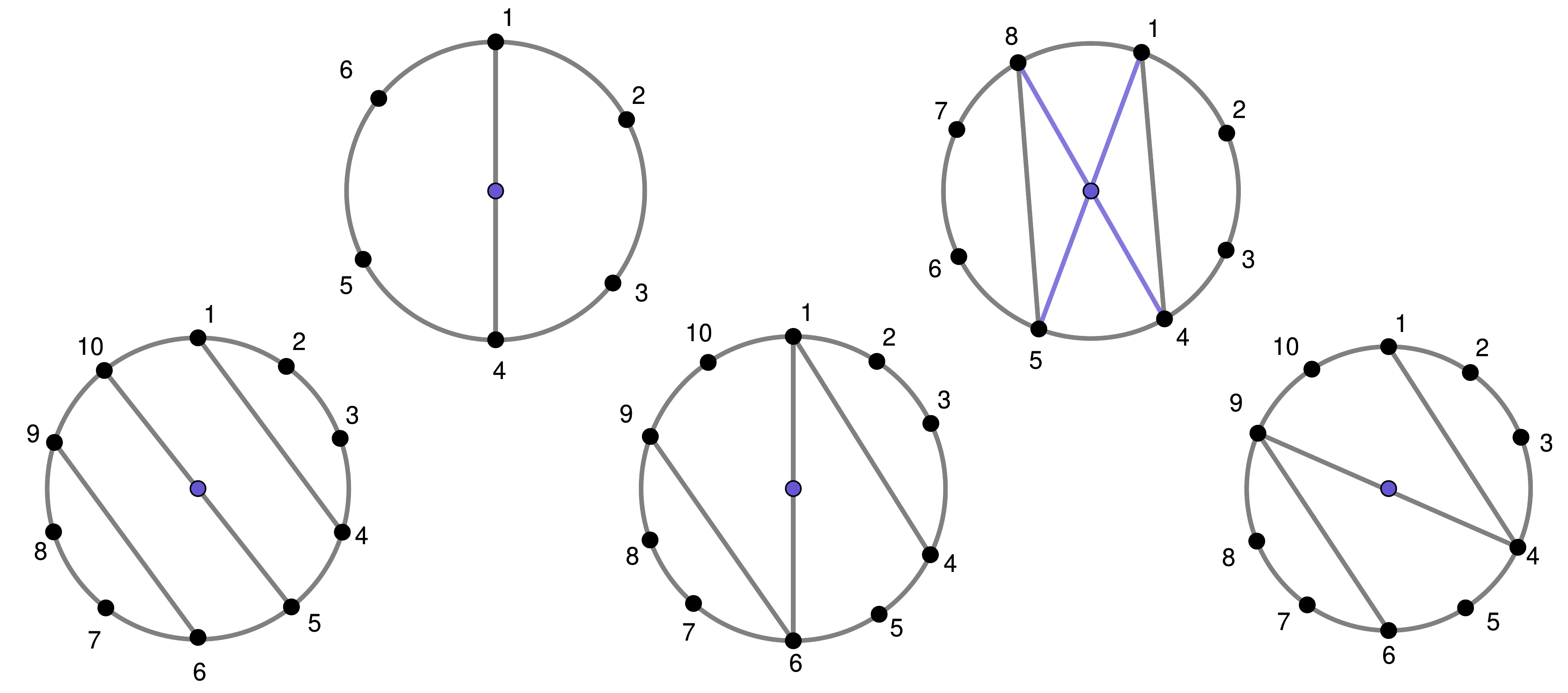} 
\setlength{\belowcaptionskip}{8pt}
\caption{ invariant quadrangulations for n=1,2,3}
\label{fig: primex} 
\end{figure}
\vskip .5cm
\noindent {\bf n=1}:  We have 3 quadrangulations $\{14,25,36 \}$ which remain invariant under rotation by  $\pi$. \bea p_1 = \frac{1}{6} (3 + 3) =1 \nonumber  \eea
{\bf n=2}: There are 4 quadrangulations $\{(1+i~ 4+i ~,5+i~8+i))\}$ with $i=0,..,3$ which remain invariant under rotation by $\pm \frac{\pi}{2}$.  \bea p_2 = \frac{1}{8} (12 + 4) =  2 \nonumber  \eea 
{\bf n=3}: There are 15 quadrangulations $\{(1+i~ 4+i, 5+i~ 10+i, 6+i~ 9+i),(1+i ~4+i, 1+i~ 6+i, 6+i~ 9+i),(1+i ~4+i, 4+i~ 9+i, 6+i~ 9+i) \}$  with $i=0,...,4$\bea p_3 = \frac{1}{10} (55 + 5+5+5) = 7  \nonumber \eea

\subsection{counting primitives for $\phi^{p}$  case } \label{primitivecountp}

\begin{figure}[h!] 
\centering 
\includegraphics [scale=0.5] {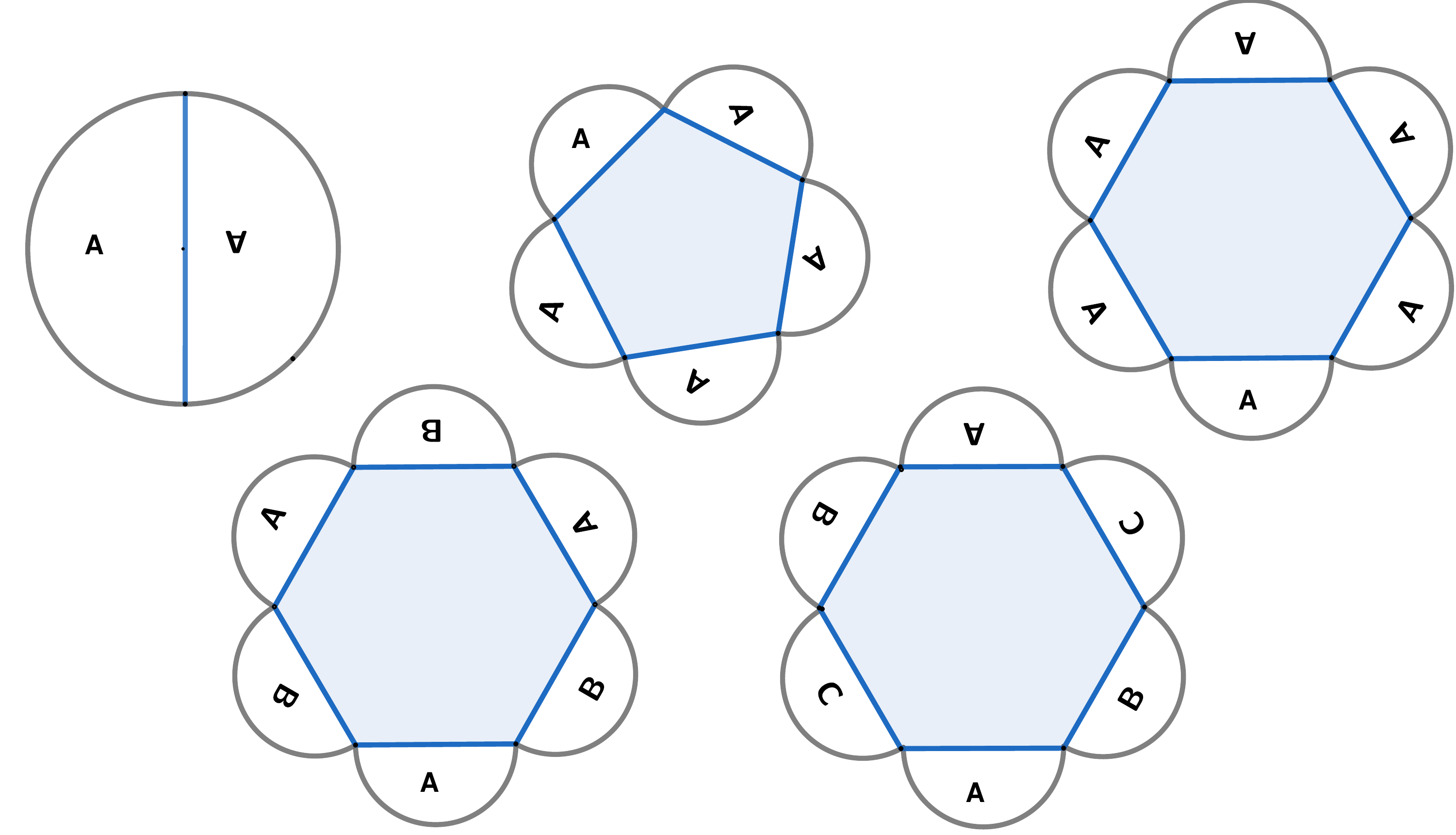} 
\setlength{\belowcaptionskip}{8pt}
\caption{The dissections invariant under some rotation for  $p=5$ it is clear that $d=1,5$ and are as shown in the first two diagrams starting clockwise from the left corner. For $p=6$ with $d=1,2,3,6$ and the invariant dissections are the ones shown in first and last three diagrams of figure(\ref{fig: genpprims})}
\label{fig: genpprims} 
\end{figure}

We shall now extend our analysis for the quartic case to any general $p$ and provide a formula for the number of primitive accordiohedra of dimension $n$.  The number of  {\it primitives p-angulations} of an $(p-2)n+p$-gon is the same as the number of orbits of the cyclic group $\mathbb{Z}_{(p-2)n+p}$ when it acts on the set of all p-angulations. There number of such orbits can be straightforwardly computed from {\it Burnside's lemma}  just as we had done in (\ref{primicountquart}). We proceed analogously to the quartic case (\ref{primicountquart}) by noting that the centre of the circle is invariant under any rotation and can lie:

 (1) On a diameter, this happens only when $n$ is odd and leaves the $p$-angulation invariant under a rotation by $\pi$  (see figure (\ref{fig: genpprims})).
 
 (2) Inside an invariant cell, in this case we have $p$-angulations for every $d~ |~ Gcd(p,n)$ which is invariant under rotation by $\frac{2 \pi}{d}$ (see figure (\ref{fig: genpprims}) ).

The total number of $p$-angulations of an $(p-2)n+p$-gon into $(n+1)$ $p$-gons is given by the Fuss Catalan number $F_{p,n} =\frac{1}{(p-2)n+p} \left(  (p-1)n \atop{n}\right)$ (see appendix \ref{appendixA} for a proof of this).

In case (1) there are $\frac{(p-2)n+p}{2}$ choices for the diameter and $F_{p, (n+1)/2 }$ choices for $A$. Thus, there are a total of $\frac{(p-2)n+p}{2}F_{p, (n+1)/2 }$ invariant $p$-angulations under a rotation by $\pi$.

In case (2) there is an  invariant cell and the remaining $n$ cells can be divided into $i=\frac{p}{d}$ parts for every $d ~|~ Gcd(p,n)$ in $ F_{k_1,p}$, $ F_{k_2,p}$,...$F_{k_{i},p} $ ways s.t. $k_1+k_2+...+k_i = \frac{n}{d}$ which we call $A$, $B$, $C$ etc. For each such $d$ there are $\phi(d)$  $p$-angulations which remain invariant under a $\frac{2\pi}{d}$ rotation, where $\phi(d)$ is the Euler totient function which counts positive integers up to $d$ which are relatively prime to it.  

 For if  $d= p_1^{a1} p_2^{a_2} ... p_r^{a_r}$ is the prime factorisation of $d$ then $\phi(d)$ is given by:

\bea
\phi(d) = d \left( 1-\frac{1}{p_1}\right) \left( 1-\frac{1}{p_2}\right)... \left( 1-\frac{1}{p_r}\right) \nonumber
\eea
Thus, the total number of such $p$-angulations once we also include identity rotation is:
\bea\label{phipprimitivecount}
p_{n} &=& 
\begin{cases}
\frac{1}{(p-2)n+p} F_{n,p}+ \frac{1}{2} F_{\frac{n+1}{2} ,p}  + \frac{1}{p}\sum_{d | Gcd(n,p)} \phi(d) \sum_{k_1+...+k_i= n/d} F_{k_1,p} F_{k_2,p}...F_{k_i,p},~~~ {\rm if}~n~{\rm is ~odd} \nonumber \\
 \frac{1}{(p-2)n+p} F_{n,p}+  \frac{1}{p}\sum_{d | Gcd(n,p)} \phi(d) \sum_{k_1+...+k_i= n/d} F_{k_1,p} F_{k_2,p}...F_{k_i,p},~~~~~~~~~~~~~~~~{\rm if}~n~{\rm is ~even} \nonumber
\end{cases} \\\\ 
p_{n} &=&
\begin{cases} 
 \frac{1}{(p-2)n+p} F_{n,p}+ \frac{1}{2} F_{\frac{n+1}{2} ,p}  + \frac{1}{p} \sum_{d | Gcd(n,p)} \phi(d) \tilde{F}_{n/d,p,p/d} , ~~{\rm if} ~n ~{\rm is ~odd} \nonumber \\
 \frac{1}{(p-2)n+p} F_{n,p}+ \frac{1}{p} \sum_{d | Gcd(n,p)} \phi(d) \tilde{F}_{n/d,p,p/d} ,~~~~~~~~~~~~~~~ ~{\rm if} ~n ~{\rm is ~even}
 \end{cases}
\eea
where, we have used the combinatorial identity 
\bea \label{combiident}
 \tilde{F}_{n/d,p,p/d} =  \sum_{k_1+...+k_i= n/d} F_{k_1,p} F_{k_2,p}...F_{k_i,p} = \frac{p/d} {(p-2)n+p} \left( (p-1)n+\frac{p}{d} \atop{n}  \right) 
\eea
with $\tilde{F}_{n,p,1}= F_{n,p}$
which we shall prove in appendix \ref{B}.

To determine the weights we would need to actually identify the primitive accordiohedra, then we need to find all the vertices of the  accordiohedra starting with these primitives as the reference $p$-angulations. There is no general classification for primitives of an arbitrary dimension $n$ to our knowledge since they grow as $p^n$. We provide a complete classification for $n \le 3$ and give the compute the corresponding weights.

\subsection{primitives and weights for $n=2$ case} \label{2case}
\begin{figure}[h] 
\centering 
\includegraphics[scale=0.3] {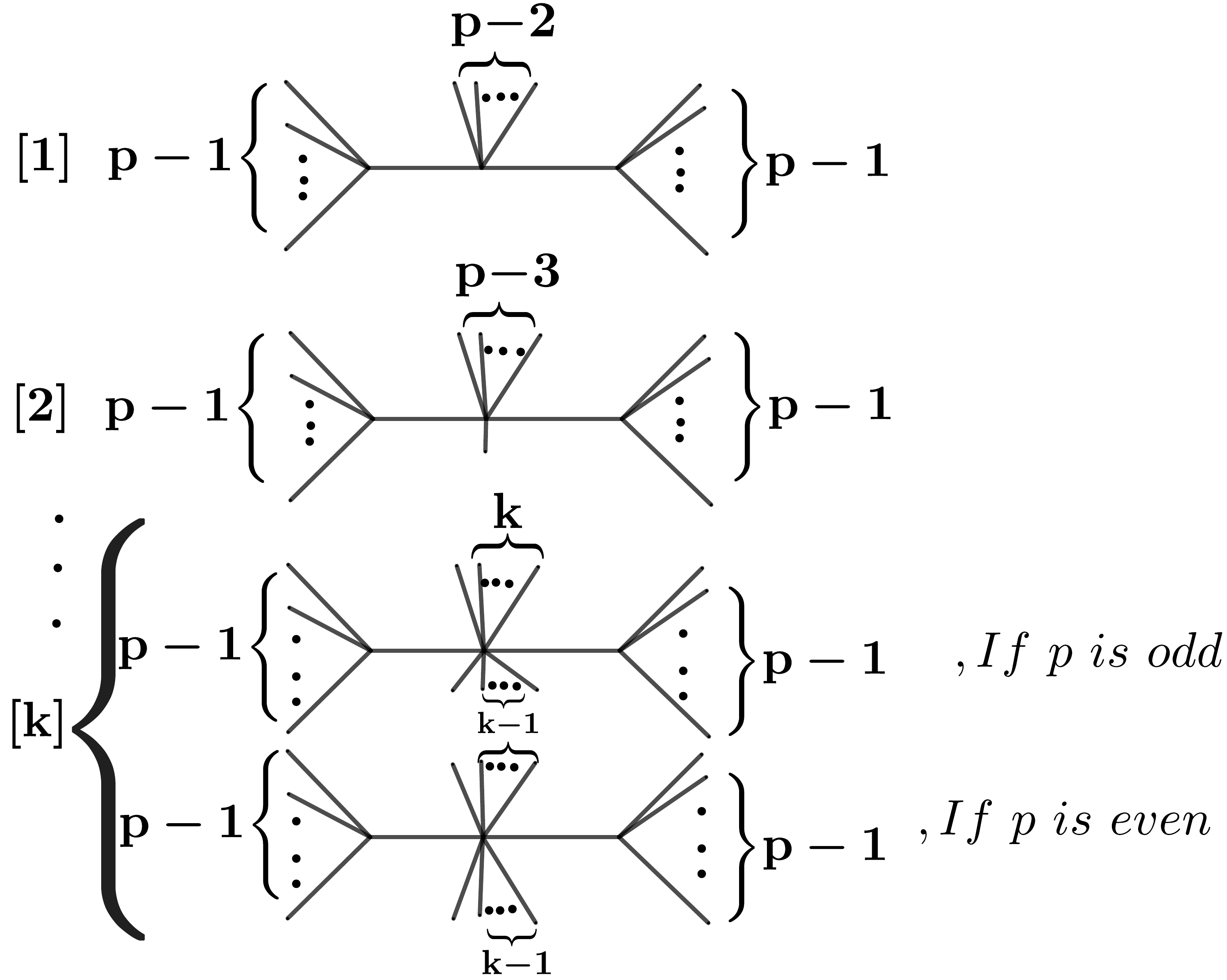}
 \setlength{\belowcaptionskip}{8pt}
\caption{The primitive graphs for n=2 case. }
\label{n=2 primitives} 
\end{figure}

We would like to provide the details of the primitives and weights for $\phi^{p}$ interactions for the $n=2$ ( 2d case). In this case there are 3 vertices with $3p-4$ legs. We could try to recursively construct these graphs  from the $n=1$ graphs.  So without loss of generality we consider only Feynman graphs in which two  of the vertices lie in a line as shown in the figure(\ref{n=2 primitives}). The 3rd vertex can then be made to lie on the central line connecting the first two vertices and it can be either above or below this line. These graphs can be denoted  as $(k_1,k_2)$ such that $k_1+k_2=p-2$, where the 3rd vertex has $k_1$ legs above and $k_2$ legs below the central line.  

Since, we are only interested in primitives the graphs for which 3rd vertices are $(k_1,k_2)$ and $(k_2,k_1)$ correspond to the same primitive graph. Thus, without loss of generality we can choose the diagrams in which the 3rd vertex has more legs above the central vertex than below it to be primitive graphs namely $(p-2,0)$, $(p-3,1)$,..., $(\left \lceil\tfrac{p-2}{2} \right  \rceil, \left \lfloor \tfrac{p-2}{2} \right  \rfloor)$. We shall call them $[1]$,$[2]$,...,$[k]$, where $k= \left \lfloor \tfrac{p}{2} \right  \rfloor$.

The primitives are all shown in the above figure(\ref{n=2 primitives}). We shall now show that these are the only primitives.
It is clear that all the graphs above are inequivalent under cyclic permutations. As explained earlier the  total number of such graphs is given by the Fuss-Catalan number 
\bea
 \frac{1}{3(p-2)+1}  \left( {3(p-1)} \atop{3} \right) = \frac{(p-1)(3p-4)}{2} \nonumber
\eea 
\vskip .2cm
We shall show that if we perform the channel sum starting with these primitives we generate all the graphs. A cyclic permutation corresponds to a clockwise rotation of the labels  and every graph returns to itself after a rotation of  period $3p-4$ but a graph for which the 3rd vertex is symmetric about the central line returns to itself after only half a rotation i.e has a period $\tfrac{3p-4}{2}$.  The only such  graph  is the last graph in the case when $p=2k$. Thus, the total number of graphs generated by performing sum over all the channels is :
\bea
\begin{cases}
k (3p-4) ~~~~~~~~~~~~~~~~~~~ {\rm if} ~p=2k+1 \\
(k-1)(3p-4)+ \frac{3p-4}{2}~~  {\rm if} ~p=2k 
\end{cases} \nonumber
\eea
using 
\bea
\begin{cases}
k = \frac{p-1}{2} ~~~ {\rm if} ~p=2k+1 \\
k=\frac{p}{2}~~~ ~~~ {\rm if} ~p=2k 
\end{cases}\nonumber
\eea
we get the total number of graphs to be $\frac{(p-1)(3p-4)}{2}$ which agrees with our results from (\ref{phipprimitivecount}).
As explained in the previous subsection the accordiohedra generated by starting with a particular graph (or p-angulation ) depends only on the relative configuration of the diagonals  and in this case since there are 2 diagonals the only possibilities are:
 \begin{enumerate}
 \item The diagonals meet as in $ [1] $, in this case the accordiohedron is an associahedron $A_n$.(see (\ref{fig: accordion2}))
 \item The diagonals do not meet as in $ [2] $,... ,$ [k]$,  in all these other cases the accordiohedron turns out to be a square.
 \end{enumerate}
We can provide a mapping between the vertices of the Stokes polytope $AC^{P}_{4,2}$ and ${\it AC^{P}_{p,2}}$ for the $n=2$ case as follows:
\begin{enumerate}
\item  When the two diagonals meet then $P= \{i ~p+i, i~ i+2p-2 \}$ with $i=1,...,3p-4$  and we could map the vertices of the Stokes polytope which is a pentagon in case with the pentagon corresponding to the accordiohedra once we notice that the all $i=1,...,8$ which is part of some diagonal $ij$ do not  appear in a vertex of the Stokes polytope. For example in the case of $P= \{(1~4, 1~6 \}$ only $i=1,3,4,5,6,8$ appear and similarly in the case of the accordiohedron exactly 6 $i =1,~p-1,~p, ~2p-3,~2p-4,~3p-4 $ out of a possible $3p-4$ appear thus we could trivially define a map between the two as follows:
\bea
i &\rightarrow & i+f(i) \nonumber\\
{\rm with} ~ f(i) &=&
\begin{cases}
1, ~~~~~~~~~ {\rm if} ~i=1\\
(p-4), ~~ {\rm if} ~ i=3,4,\\
2(p-4), ~ {\rm if} ~ i=5,6\\
3(p-4), ~ {\rm if} ~ i=8\nonumber\\
\end{cases}
\eea

\item When the two diagonals do not meet then we notice that there are several possible choices of the diagonals for $p>5$, but we notice that the maximum number of $i$'s which can appear for any choice of $P$ is 8, since each diagonal appears twice thus there are only 4 possible $ij$'s. We can thus identify these $i$'s with $i=1,...,8$ and define a mapping.
\end{enumerate}
 For example we provide such a mapping between the Stokes polytope corresponding to $Q=\{(14, 58)\}$ and the accordiohedra corresponding to $p=5,6$ (see (\ref{fig: accordion2})  )below:
 \vskip .2cm
 
{\bf p=5}:  With $P=\{(15, 610) \}$\bea f(i) = \begin{cases} 1, ~~~~~~~~~{\rm if} ~i=1\\
(p-4), ~{\rm if}~ i=2,3,4,5\\
2(p-4), ~{\rm if}~ i=6\\
3(p-4), ~{\rm if}~ i=7,8\nonumber  \end{cases} \eea
{\bf p=6}:  With $P=\{(16, 914) \}$\bea f(i) = \begin{cases} 1 ~~~~~~~,~~{\rm if} ~i=1\\
(p-3), ~{\rm if}~ i=2,3,4,5,6\\
2(p-3), ~{\rm if}~ i=7,8 \nonumber  \end{cases} \eea
It is straightforward to find such a mapping for general $p>4$  and a general choice of $P$. \footnote{It is a well known fact that there is a unique 2d convex polytope with a given number of vertices, so the above mapping is not really needed here but in the case of 3 and higher dimensional polytopes ,there could be several polytopes with same $f$-vector, thus we would need such a mapping to be sure that the polytopes are isomorphic.}\\

\subsection{primitives for $n=3$ case} \label{3case}
We would now try to find all primitive graphs for the $n=3$ case. In this case there are 4 vertices with $4p-6$ legs. As before we can try to recursively construct primitive graphs from $n=2$ case. 

\begin{figure}[!ht] 
\centering 
\includegraphics[scale=0.35] {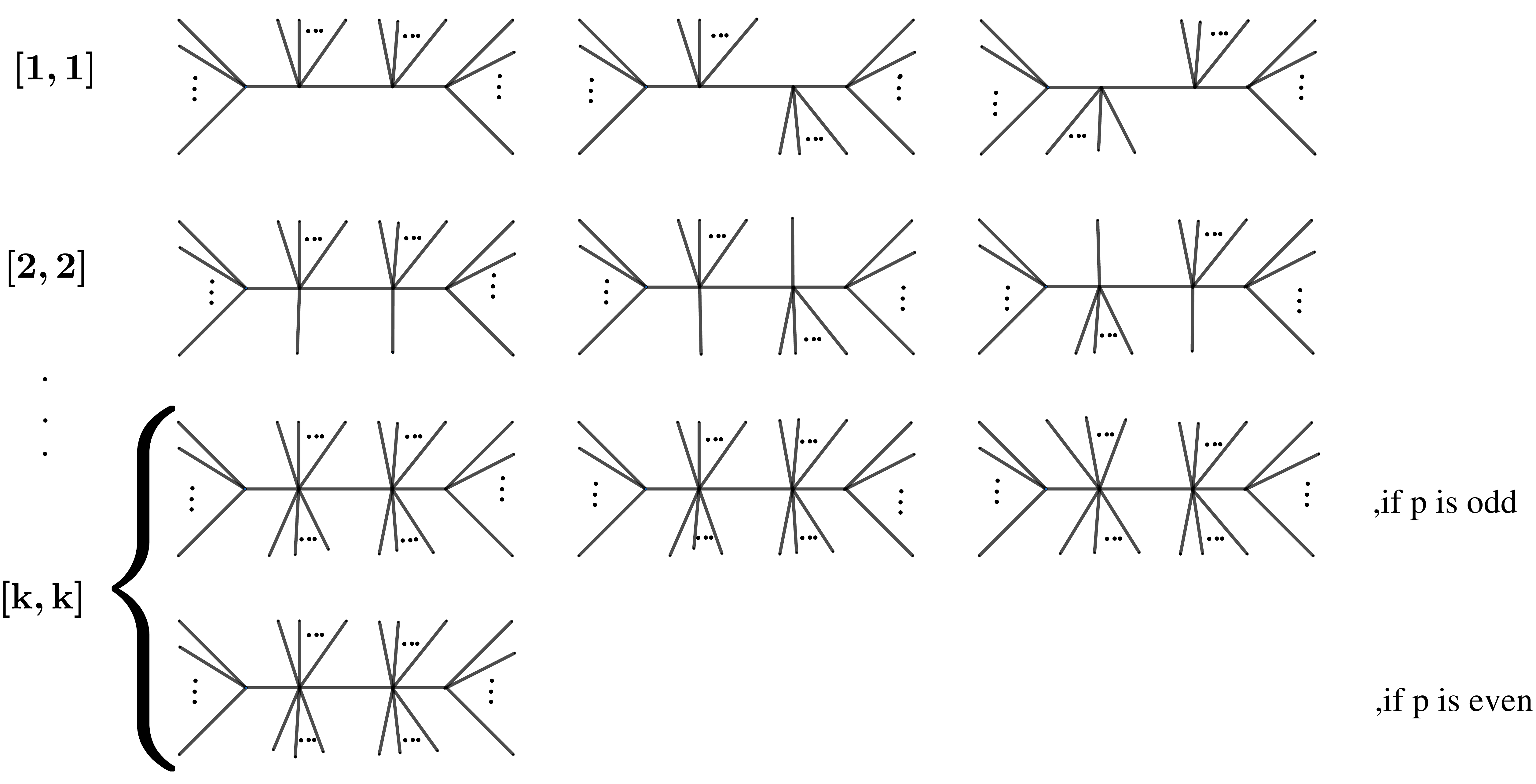}
 \setlength{\belowcaptionskip}{8pt}
\caption{The primitive graphs of the type $[i,i]$. }
\label{n=3 primitives1} 
\end{figure}
There are two possible ways we add the 4th vertex which could be any of the $k= \left \lfloor \frac{p}{2}\right \rfloor$  types in the $n=2$ case :
\begin{enumerate}

\item We can add another central vertex  either above or below the central line to a $n=2$ graph. We call these graphs $[i,i]$ and $[i,j]$. ( see  figures (\ref{n=3 primitives1}), (\ref{n=3 primitives2})).

\item We can add the vertex to any one of the external legs of the 3rd vertex. We denote these graphs by $(k_1, k_2, k_3)$, where $k_1+ k_2 + k_3 =p-3$.

\end{enumerate}
There are 3 primitives of the type $[i,i]$ for each $i=1,\cdots, k$ since the graphs where both vertices are down are just cyclic permutations of the graph with both vertices up. In case where $p$ is even you also have a vertex with equal number of legs above and below the central line, thus there is only one such primitive corresponding to this case. The graphs with one central vertex up and the other down (the 2nd and 3rd graphs for each  $[i,i]$  ) have half periods i.e. under cyclic permutations they go back to themselves after $2p-3$ operations. The same is also true for the symmetric vertex when $p$ is even. All other graphs have full period of $4p-6$.

\begin{figure}[h] 
\centering 
\includegraphics[scale=0.46] {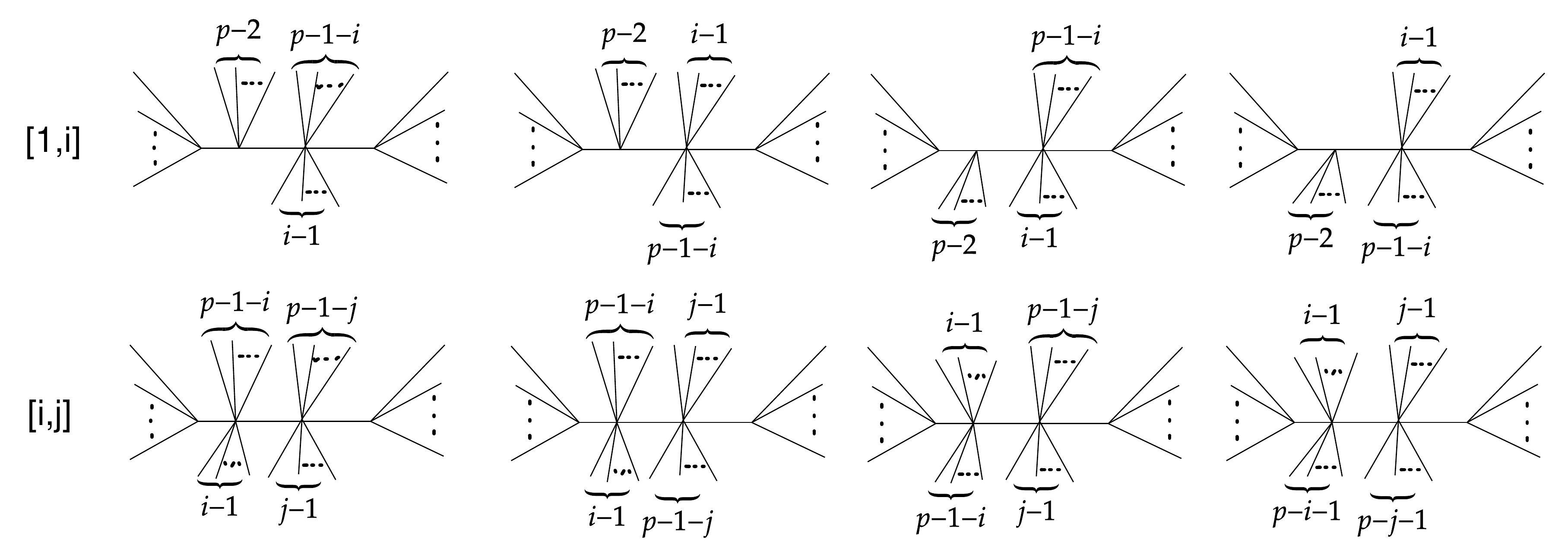}
 \setlength{\belowcaptionskip}{8pt}
\caption{The primitive graphs of the type $[i,j]$.  }
\label{n=3 primitives2} 
\end{figure}
There are 4 primitives for each  $[i,j]$ since now the graphs with both vertices down are inequivalent to the ones with both vertices up under cyclic permutations. When $p$ is even we have only 2 primitives of the type  $[i,k]$  since $[k]$ is symmetric. All these graphs all have a period of $4p-6$.

These possibilities are summarised in the table \ref{tab:table1} below:

\begin{table}[ht]
\label{tab:table1}
     \begin{center} 
   \scalebox{0.8 }{  \begin{tabular}{| c | c | c |} 
    \hline
    Type of primitive  & number of primitives & period of the primitive  \\
    \hline 
    & & \\
    $ [i,i]$ &  3 & ~~1 with $4p-6$   \\ 
    with $ i=1,..., k-1 $ & & 2 with $\tfrac{4p-6}{2}$ \\
    \hline 
   & &  \\
    $[k,k]$ & 1 , if $p$ is even &  $\tfrac{4p-6}{2}$\\
    & &  \\
                                                & 3 ,  if $p$ is odd     &  ~~1 with $4p-6$   \\
                                                &                                   & 2 with $\tfrac{4p-6}{2}$ \\
 \hline
 & & \\ 
          $[i,j]$ & &  \\  
          with $i,j =1,...,k-1$ & 4  & $ 4p-6 $ \\
          $i\neq j$ & & \\                                    
 \hline 
 & & \\ 
          $[i,k] $ & 2 , if p even  & $4p-6$  \\  
          with $i =1,...,k-1$ &  4 , if p odd & $ 4p-6 $ \\
           & & \\                                    
 \hline 
    \end{tabular}}
     \caption{Primitive graphs of type 1.}
  \end{center}
\end{table} 

We could also consider graphs of the type $(k_1, k_2, k_3)$ such that $k_1+k_2+k_3 =  p-3 $. Since there are $\left(n-1 \atop{r-1} \right)$ non zero solutions of $x_1+...+x_r =n$, in this case we have the following possibilities: 
\begin{itemize}
\item We have a graph of the type (0,0,p-3) with period $4p-6$.
\item We can have a  two graphs  $(k_1,0,k_3)$ and $(0, k_1, k_3)$ for each $k_1+k_3 = p-3$ with $k_1 \neq k_3$ (which are inequivalent since they are reflections of each other) and when $p$ is odd we also have one graph  $(k_1,0,k_3)$ with $k_1=k_3$ with period $4p-6$. Thus, there are \\
\bea
\begin{cases}
 \left(p-4\atop{1} \right)  ~~~~= (p-4) , ~{\rm if}~ p ~{\rm is ~odd} \\
 \left(p-4\atop{1} \right)-1 = (p-5), ~{\rm if}~ p ~ {\rm is ~even} \nonumber
\end{cases}
\eea
such diagrams.
\item We have one graph for each $(k_1,k_2,k_3)$ with $k_1 =k_2 \neq k_3$ with period $4p-6$. In this case we have 
\bea
\begin{cases}
 \lfloor \frac{p-3}{2} \rfloor,~~~~~~{\rm if}~ p \neq 3k \\
 \lfloor \frac{p-3}{2} \rfloor-1, ~{\rm if}~ p = 3k\nonumber
\end{cases}
\eea
such diagrams.
\item When $p=3k$ we have exactly one graph with $k_1=k_2=k_3$ which has a period $\frac{4p-6}{3}$.
\item We have two graphs $(k_1,k_2,k_3)$ and $(k_2,k_1,k_3)$ for each $k_1\neq k_2 \neq k_3$.  In this case we have 
\bea
\begin{cases}
\frac{\frac{(p-4)(p-5)}{2}- 3 \lfloor \frac{p-3}{2} \rfloor}{3} ,~~~~~ ~~{\rm if}~ p \neq 3k \\
\frac{\frac{(p-4)(p-5)}{6}- 3\left (\lfloor \frac{p-3}{2} \rfloor-1\right)-1}{3}, ~~{\rm if}~ p = 3k \nonumber
\end{cases}
\eea
\end{itemize}
These possibilities are summarised in the table 2  below.
\begin{table}[t!] 
\label{tab:table2}
  \begin{center}
  \scalebox{0.8}{
   \begin{tabular}{| c | c | c |} 
   
    \hline
    Type of primitive  & number of primitives & period of the primitive  \\
    \hline 
    & & \\
    $ (0,0,p-3) $ &  1 &   $4p-6$  \\ 
    \hline 
   & &  \\
    $ (k_1, 0 ,k_3 )$ & 1 , if $p$ is odd &  $4p-6$\\
   $k_1 =k_3 =\frac{p-3}{2}$ &0 , if $p$ is even  &  \\
        \hline 
           & & \\
         $ (k_1, 0 ,k_3 )$ & $p-4$ , if p is even &  $4p-6$  \\
   $k_1 \neq k_3 $ &  $p-5$ , if p is odd &  \\
   \hline 
    & & \\
         $ (k_1, k_2 ,k_3 )$ & $\tfrac{p-4}{2}$ , if p is even &  $4p-6$  \\
         & ~~~~~~~~~~$p\neq 3k$ &   \\
   $k_1 =  k_2 $ &  $\tfrac{p-5}{2} $ , if p is odd & $4p-6$  \\
     & ~~~~~~~~~~$p\neq 3k$ &   \\
     &  $\lfloor \tfrac{p-3}{2}\rfloor -1 $ , if $p =3k$ &   $4p-6$  \\
   \hline 
    & & \\
         $ (k_1, k_2 ,k_3 )$ & 1 , if $p=3k$ &  $\tfrac{4p-6}{3}$  \\
   $k_1 =  k_2=k_3 $ & 0 , otherwise &  \\
   \hline 
    & & \\
         $ (k_1, k_2 ,k_3 )$ &  $ \tfrac{(p-4)(p-5)}{6} - \tfrac{p-4}{2}$ , if $p$ is even & $4p-6$    \\
         $k_1 \neq k_2\neq k_3 $  & ~~~~~~~~~~~~~~~~~~~~~~~~~~ $p\neq 3k $ &   \\ 
&$ \tfrac{(p-4)(p-5)}{6} - \tfrac{p-5}{2}$ , if $p$ is odd & $4p-6$    \\
         & ~~~~~~~~~~~~~~~~~~~~~~~~~~ $p\neq 3k $ &   \\ 
   & $ \tfrac{(p-4)(p-5)}{6} - \lfloor\tfrac{p-3}{2} \rfloor +\tfrac{2}{3}$ , if $p= 3 k $   & $4p-6$    \\
 \hline   
       \end{tabular}}
  \caption{Primitive graphs of type 2.} 
  \end{center}
\end{table} 

We can now find the total number of $p$-angulations by summing over all channels by multiplying columns two and three of the tables above and adding them up. 
\bea
\text{number of p-angulations} = \sum_{\sigma} (\text{ primitive  } \sigma ) \times (\text {period of } \sigma ) \nonumber
\eea

The result of this exercise turns out to be $\frac{(p-1)(2p-3)(4p-5)}{3}$ which matches with the expected Fuss-Catalan number which agrees with equation (\ref{phipprimitivecount})
\bea
\frac{1}{4(p-2)+1}  \left( {4(p-1)} \atop{4} \right) = \frac{(p-1)(2p-3)(4p-5)}{3} \nonumber
\eea
\begin{figure}[h!] 
\centering 
\includegraphics[scale=0.35] {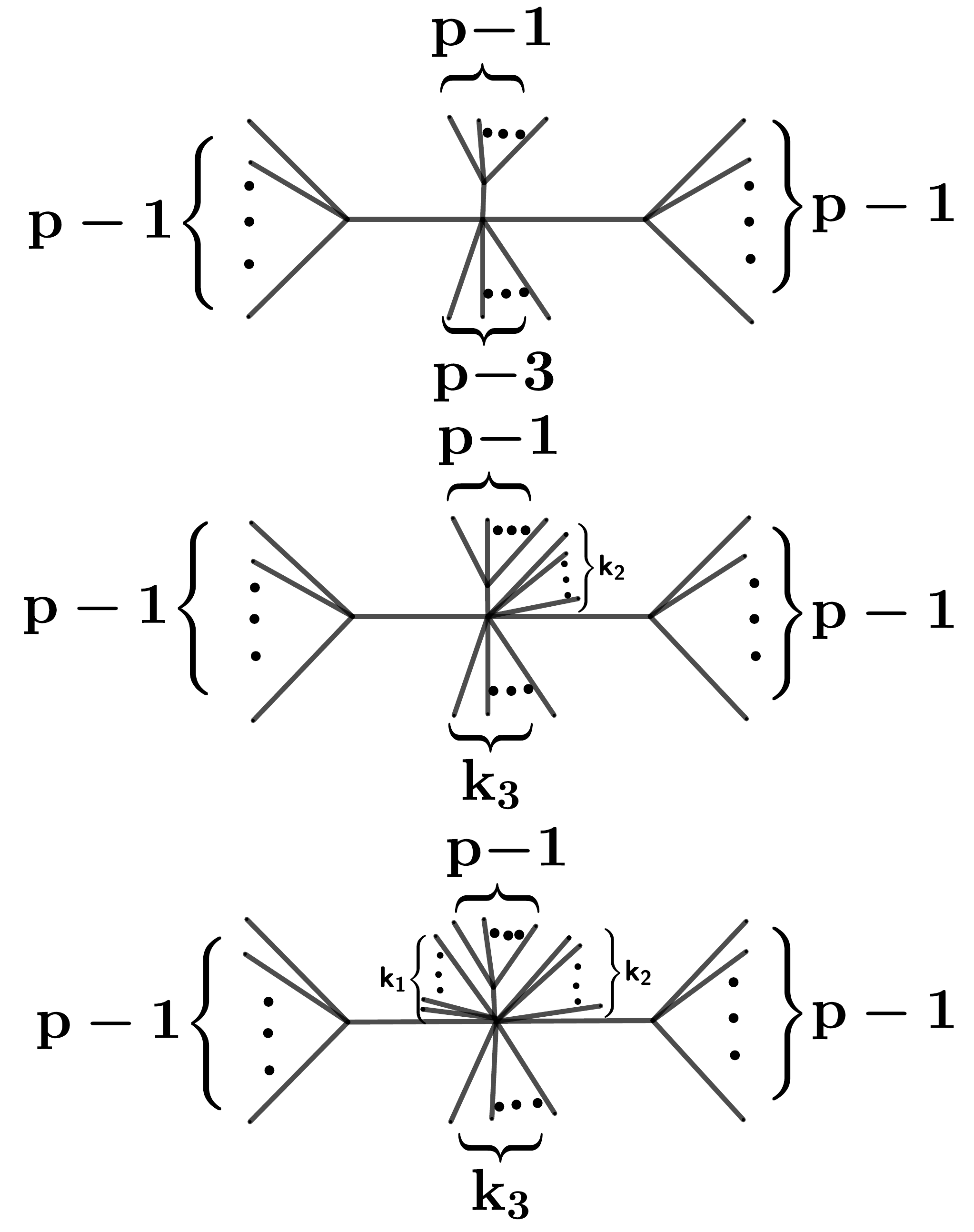}
 \setlength{\belowcaptionskip}{10pt}
\caption{The primitive graphs of the type $(k_1,k_2,k_3)$.}
\label{n=3 primitives2} 
\end{figure}

Since there are 3 diagonals now the relative configuration of diagonals can be of one of the following types:
\begin{enumerate}
\item None of the diagonals meet - in this case the corresponding accordiohedron is a cube. There are   $1+ \left \lfloor \frac{2 (p-3)^2}{3} \right \rfloor$ graphs of this type  namely $[i,i]$, $[i,j]$ with $i,j =2,...,k$ and $(k_1,k_2,k_3) $ with $ k_1, k_2,k_3 \neq 0$.
 \item Two of  the  diagonals meet - in this case the corresponding accordiohedron is of the mixed type. There are $3p-10$ graphs of this type namely  $[1,j]$ with $j=2,...,k$ and $(k_1,0,k_3) $ with $ k_1, k_3 \neq 0$.
\item All three diagonals meet at a vertex or form zig-zag configuration in this case the corresponding accordiohedron is an associahedron. There are $3$ graphs of this type namely  $[1, 1]$.
\item  All three diagonals meet and form an inverted {\it U} configuration in this case the corresponding accordiohedron is of the Lucas type. There is exactly one graph of this type which is $(0,0,p-3)$.
\end{enumerate}
Thus the  total number of primitives is :
\bea
3p-5 + \left \lfloor \frac{2 (p-3)^2}{3} \right \rfloor =  \left \lceil  \frac{(p-1)(2p-1)}{3}\right \rceil \nonumber 
\eea 
which agrees with we our general formula (\ref{phipprimitivecount}).

The accordiohedra for $n=3$ we get continue to be one of the four kinds of Stokes polytopes. We could define a function from vertices of the Stokes polytopes to that of the accordiohedra as we had done in the $n=2$ case to establish that this is indeed the case. We can thus continue to use the same names  Lucas, Mixed etc for the $n=3$ Stokes polytopes for accordiohedra as well. We expect that at sufficiently higher $n$, accordiohedra will be generated which do not correspond to any Stokes polytope.

\subsection{Determination of the weights} \label{weights}
In this section we shall provide a simple method to determine the weights for the general case and demonstrate the method in a few examples. We recall that we had the reduced amplitude $\tilde{{\mathcal M_n}} $ which is a weighted sum of canonical forms of all the primitive accordiohedra of a given dimension $n$. We would like to determine the weights such that this gives the full amplitude i.e. $ \tilde{{\mathcal M_n}} ={\mathcal M_n}$.

The full amplitude $ {\mathcal M_n}$ is given by : 

\bea
{\mathcal M_n} = \sum_{all~ i_k j_k} \prod_{k=1}^{n}\frac{1}{X_{i_k j_k}}  \nonumber
\eea
where, the sum is over all $( i_1 j_1,...,i_n j_n )$ that form a complete $p$-angulation.

Thus, to get the full amplitude from the partial amplitude we need to impose the constraint that each $\prod_{k=1}^{n} \frac{1}{X_{i_k j_k}}$ appears exactly once. 

But as we had emphasised before the accordiohedron depends only on the relative configuration of diagonals of the reference $p$-angulation which does not change under rotations and thus it is sufficient to impose these constraints for the primitive $p$-angulations.
\bea
\sum_{i=1}^{l} n_{p}^{i}\alpha_{p}^{i} =1  ~{\rm for ~each~ primitive}~ 1 \le i \le l \nonumber
\eea
where, $n_{p}^{i}$ is number of times primitive $i$ appears in the vertices of all accodiohedra, $\alpha_{p}^{i}$ are the corresponding weights.

Since, we have managed to classify all the primitives unto $n=3$ we should be able to implement this straightforward procedure to get all the weights and we shall now discuss our results.

We shall first see what these conditions are for $n=2$ in the $p=5,6$ cases.

${\bf p=5}$ : In this case there are two primitives as we had explained in the section (\ref{2case}) and we get:

\bea
3 \alpha_{5}^{1} +\alpha_{5}^{2} &=&1\nonumber \\
 \alpha_{5}^{1} +4 \alpha_{5}^{2} &=&1 \nonumber
 \eea
 which can be solved to give:
 
\hspace*{180pt} $ \alpha_{5}^{1} =\frac{3}{11} $, $\alpha_{5}^{2} =\frac{2}{11} $

${\bf p=6}$ : In this case there are 3 primitives and we get:
\bea
2 \alpha_{6}^{1} +\alpha_{6}^{2} + 2 \alpha_{6}^{3}&=&1\nonumber \\  \alpha_{6}^{1} + 4 \alpha_{6}^{2} &=&1 \nonumber \\ 
 \alpha_{6}^{1} +2 \alpha_{6}^{3}&=&1 \nonumber
  \eea 
which can be solved to give:

\hspace*{180pt} $ \alpha_{6}^{1} = \frac{1}{3}$ $\alpha_{6}^{2}=\frac{1}{6}$ , $ \alpha_{6}^{3}=\frac{1}{3} $

We can similarly do this for any $p$ with $n=2$ and the results are the following :

For $p=2k$
\bea
\alpha_{(p-2-i,i)}=
\begin{cases}
\frac{1}{6}~,i ~ {\rm even} \\ 
\frac{1}{3}~,i~ {\rm odd}
\end{cases} \nonumber 
\eea

and For $p=2k+1$
\bea
\alpha_{(p-2-i,i)}= \frac{k+1+i}{3p-4}\nonumber 
\eea
with $i=0,..., k-1 $.

The $\alpha$'s for $n=3$ case with $p\le 12$ are given below (for the sake of brevity we shall call $\alpha$'s corresponding to $[i,j], (k_1,k_2,k_3)$ as $[i,j], (k_1,k_2,k_3)$):
\vskip .2cm
\noindent{\bf If $p$ is even then} : 
\vskip .2cm

\noindent $[i,i] = \tfrac{1}{24},\tfrac{5}{24},\tfrac{1}{24},... $ ; $ [1,i]  = \tfrac{3}{24},\tfrac{1}{24},\tfrac{3}{24},... $ ; $ [2,i] = \tfrac{3}{24},\tfrac{5}{24},\tfrac{3}{24},... $ ; $ [3,i]  = \tfrac{3}{24},\tfrac{1}{24},\tfrac{3}{24},... $ ; ... 
\\ $ (k_1,k_2,k_3) = (k_2,k_1,k_3 )=\tfrac{6}{24},\tfrac{2}{24},\tfrac{6}{24},...$; $ (k_1,0,k_2) = (0,k_1,k_2 )=\tfrac{2}{24}$ ,  $ (0,0,p-3) = \tfrac{2}{24}$ \\

\noindent{\bf If $p$ is odd  then the results for the first few cases are} :
\vskip .2cm
\noindent{\bf p=5} :  $[i,i] = \tfrac{1}{20}, \tfrac{3}{20}$ with $i=1,2$; $ [1,2] = \tfrac{2}{20}$; $ (1,1,0)= \tfrac{2}{20}$; $ (0,0,2)= \tfrac{2}{20}$. \\\\
{\bf p=7} : $[i,i] = \tfrac{3}{64}, \tfrac{11}{64}, \tfrac{7}{64}$ with $i=1,2,3$; ~ $ [1,j] = \tfrac{7}{64}, \tfrac{5}{64}$, with $j=2,3$ ; ~   $ [2,3] = \tfrac{9}{64}$;~ $ (1,1,2)= \tfrac{10}{64}$, \\ $ (0,1,3)=(1,0,3)= \tfrac{6}{64}$;~$ (2,0,2)= \tfrac{6}{64}$;~ $ (0,0,4)= \tfrac{6}{64}$ .\\\\
{\bf p=9} : $[i,i]= \tfrac{2}{44}, \tfrac{8}{44}, \tfrac{4}{44},\tfrac{6}{44},$ with $i=1,2,3,4$;~  $[1,j] = \tfrac{5}{44}, \tfrac{3}{44},\tfrac{4}{44}$, with $j=2,3,4$ ;~   $ [2,j]= \tfrac{6}{44},\tfrac{7}{44}$, with $j=3,4$;~$ [3,4] = \tfrac{5}{44}$;~ $ (2,2,2)= \tfrac{4}{44}$;~ $ (1,1,4)= \tfrac{8}{44}$;~$ (1,2,3)= (2,1,3)=\tfrac{6}{44}$ ;~$ (3,0,3)= \tfrac{4}{44}$,\\ $ (1,0,5)=(0,1,5)= \tfrac{6}{64}$;~ $ (2,0,4)=(0,2,4)= \tfrac{4}{44}$;~$ (0,0,6)= \tfrac{4}{44}$ .\\\\
{\bf p=11}: $[i,i] = \tfrac{10}{112}, \tfrac{21}{112}, \tfrac{9}{112},\tfrac{17}{112},\tfrac{13}{112}$ with $i=1,2,3,4,5$;~  $ [1,j] = \tfrac{13}{112}, \tfrac{7}{112}, \tfrac{11}{112},\tfrac{9}{112}$, with $j=2,3,4,5$;~ \\  $ [2,j] = \tfrac{15}{112}, \tfrac{19}{112}, \tfrac{17}{112}$, with $j=3,4,5$;~$[3,j] = \tfrac{13}{112}, \tfrac{11}{112}$, with $j=4,5$;~ $ [4,5] = \tfrac{11}{112}$;~ $ (1,1,6)= \tfrac{22}{112}$;~ \\ $ (2,2,4)= \tfrac{10}{112}$;~$ (3,3,2)= \tfrac{14}{112}$;~ $ (0,4,4)= \tfrac{10}{112}$;~ $ (1,0,7)=(0,1,7)= \tfrac{10}{112}$;~ $ (2,0,6)=(0,2,6)= \tfrac{10}{112}$;~\\ $ (3,0,5)=(0,3,5)= \tfrac{10}{112}$;~ $ (1,2,5)=(2,1,5)= \tfrac{14}{112}$;~$ (1,3,4)=(3,1,4)= \tfrac{14}{112}$;~$ (0,0,8)= \tfrac{5}{112}$ .\\\\

\section{Factorization}

One of the remarkable consequences of relating tree level scattering amplitudes to  positive geometries like associahedron, Stokes polytope is the fact that geometric factorization implied physical factorization of scattering amplitude. This in turn implied that tree-level unitarity and locality are emergent properties of the positive geometry \cite{Arkani-Hamed:2017mur, Banerjee:2018tun}. In this section we will try to argue that this is indeed the case even for planar amplitudes in  massless $\phi^{p}$ theory. We shall first argue that the geometric factorisation of accordiohedron holds and then show how this leads to the factorisation of the amplitude. 

In \cite{Banerjee:2018tun} it was shown how the factorization  of Stokes polytope leads to a recursions relation on $\alpha$'s. We shall see that even for more general Accordiohedra $\alpha$'s are required to satisfy analogous recursion relations.

Our first assertion is the following. Given any diagonal $(ij)$,  consider \emph{all} $P$ which contains $ij$ and the consider all the corresponding kinematic accordiohedron ${\mathcal AC}_{p,n}^{P}$. We contend that for each accordiohedron, the corresponding facet $X_{ij}\ =\ 0$ is a product of lower dimensional accordiohedra.  
\begin{equation}\label{factor1}
{\mathcal AC}_{p,n}^{P}\bigg|_{X_{ij}\ =\ 0}\ \equiv\ {\mathcal AC}_{p,m}^{P_{1}}\ \times\ {\mathcal AC}_{p,n+2-m}^{P_{2}}
\end{equation}
where $P_{1}$ and $P_{2}$ are such that $P_{1}\ \cup P_{2}\ \cup\ (ij)\ =\ P$. 

$P_{1}$ is the $p$-angulation of the polygon $\{ i,\ i+1,\dots,\ j\}$  and $P_{2}$ is the $p$-angulation of $\{j,j+1,\dots,n,\dots,i\}$. 
Now we know that, on ${\mathcal AC}_{p,n}^{P}$ any planar scattering variable $X_{kl}$ is a linear combination of $X_{ij}$ and remaining $X$'s which constitute $P$. Hence in order to prove this assertion we need to show that  any $X_{kl}$ with $ i \leq k  <  l  \leq j$ can be written as a linear combination of $X_{ij}$ and elements of $P_{1}$ and  similarly any variable in the complimentary set can be written in terms of $X_{ij}$ and elements of $P_{2}$.

However this is immediate since we know from the factorization property of associahedron proven in \cite{Arkani-Hamed:2017mur} that any $X_{kl} = \displaystyle \ X_{ij}\ + \sum_{i < m < n < j} X_{mn}$, some of these $X_{mn}\ \in\ P_{1}$ and the others are constrained via $X_{mn} = d_{mn}$. This proves our assertion. Thus $X_{ij} = 0$ facet factorizes into two lower dimensional accordiohedra. 

Our second assertion is that the geometric factorization implies amplitude factorization of $\phi^{p}$ theory. This assertion is based on the following two facts (For details, we refer the reader to appendix A of \cite{Arkani-Hamed:2017mur} and \cite{Arkani-Hamed:2017tmz}).
\vskip .2cm
\noindent
(1) As the accordiohedra  is a positive geometry, we know that it's canonical form satisfies the following properties satisfed by canonical form on any positive geometry ${\cal A}$  

\begin{equation}
\textrm{Res}_{H}\omega_{{\cal A}}\ =\ \omega_{{\cal B}}
\end{equation}
where we think of $\omega_{{\cal A}}$ as defined on the embedding space and $H$ is any subspace in the embedding space which contains the face ${\cal B}$.
\vskip .2cm
\noindent
(2) It is also known that if ${\cal B}\ =\ {\cal B}_{1}\times{\cal B}_{2}$ then

\begin{equation}
\omega({\cal B})\ =\ \omega({\cal B}_{1})\ \wedge\ \omega({\cal B}_{2})
\end{equation}

Thus we immediately see that 

\begin{equation}
\textrm{Res}_{ X_{ij}\ =\ 0} \;\; \omega({\mathcal AC}_{p,n}^{P})\ =\ \omega_{m}^{P_{1}}\ \wedge\ \omega_{n+2-m}^{P_{2}}\ \forall\ P. 
\end{equation}
where $m\ =\ j - i + 1$. 

We thus see that residue over each accordiohedron which contains a boundary $X_{ij}\ \rightarrow\ 0$ factorizes into residues over lower dimensional accordiohedra. 
This factorization property naturally implies factorization of amplitudes as follows. Consider the $n$-gon with a diagonal $(ij)$ (with $i,j$ such that this diagonal can be part of a $p$-angulation). This diagonal subdivides the $n$-gon into a two polygons with vertices $\{i,\dots,j\}$ and $\{j,\dots,n,1,\dots i\}$ respectively.  By considering all the kinematic accordiohedra associated to these polygons, we can evaluate  $\widetilde{M}_{\vert j - i + 1\vert},\ \widetilde{M}_{n+2 - (\vert j - i + 1\vert)}$ which correspond to left and right sub-amplitudes respectively. This immediately implies that 

\begin{equation} \label{factor2}
\widetilde{{\cal M}}_{n}\vert_{X_{ij}\ =\ 0}\ =\ \widetilde{{\cal M}}_{L}\frac{1}{X_{ij}}\widetilde{{\cal M}}_{R}
\end{equation}

This proves physical factorization. We also note that, eqns. (\ref{factor1}) and (\ref{factor2}) imply following constraints on $\alpha$'s.

\begin{equation} \label{factor3}
\sum_{P\ \textrm{containing} (ij)} \alpha_{P}\ =\ \sum_{P_{L}, P_{R}}\alpha_{P_{L}}\alpha_{P_{R}}
\end{equation}
The left hand side of the above equation involves sum over all accordiohedra ${\mathcal AC}_{p,n} ^{P}$ for  which $(ij) \in ~P$ and the right hand side involves sum over  $P_{L}$ and $P_{R}$ which range over all the $p$-angulations of the two polygons to the left and right of the diagonal $(ij)$ respectively. 
 
 It can be verified that in all the examples up to $p=12$ and $n=3$ the $\alpha_{P}$'s do indeed satisfy these constraints.
 
 For $n=1$ there is only one diagonal which we can to be $(1,p)$. The accordiohedra is always a line as we had emphasised in section (3.3) and $(1,p)$ appears in the vertex of exactly two of these lattices namely $\{1~p, ~p-1~ 2p-2 \}$ and $ \{p+1~2p+1, ~1~ p \}$. There is only way to divide $P$ into $P_L$ and $P_R$ and both these are trivial have $\alpha =1$. Thus, the above equation (\ref{factor3}) gives :
 
 \bea
 2 \alpha_P&=&1 \nonumber\\
 \alpha_P&=& \frac{1}{2}~ {\rm for~ all} ~P \nonumber
 \eea 
 
We could expect  that the set of equations (\ref{factor3}) would help in determining all the weights \cite{Banerjee:2018tun}. But, as we shall now show this is not the case as  (\ref{factor3}) provides \emph {too few} equations. In other words the  set of equations (\ref{factor3}) provide a set of necessary but not sufficient conditions.

Let, us consider the case $n=2$ for $p=5$ and the diagonal $(15)$, in this case  the $11$-gon gets divided into a $5$-gon and $8$-gon and we have weights $\alpha_{P_L} =\frac{1}{2}$ and $\alpha_{P_R} =1$. There are exactly $4$ squares and $5$ pentagons which contain the diagonal $(15)$ in their vertices. Thus, we have
\bea \label{eqn}
4 \alpha_{5}^{1} +5 \alpha_{5}^{2} = 2
\eea
We can check that for any other choice of diagonal $(ij)$ we get the same equation. It is clear that this is not sufficient to solve for $\alpha_{5}^{1}, \alpha_{5}^{2}$. The solution we had obtained using our prescription in section (\ref{weights})  namely $\alpha_{5}^{1}=\tfrac{3}{11}, \alpha_{5}^{2}=\tfrac{2}{11}$ does indeed satisfy (\ref{eqn}).

\section{Conclusions and Future work}

Re-formulating scattering amplitudes as differential forms on positive geometries (succinctly called the Amplituhedron program) has had profound impact on how we understand Quantum field theories and how properties like unitarity and locality are a natural consequence of the positive geometries.  In theories like ${\cal N}=4$ Super Yang Mills theory, the Amplituhedron program offers conceptual as well as striking technical advancements in the understanding of planar S-matrix. In the non super-symmetric world, these ideas were extended to bi-adjoint scalar theory in \cite{Arkani-Hamed:2017mur} where it was shown that the corresponding amplituhedron is an associahedron in Kinematic space and the canonical form on this associahedron was proportional to the scattering amplitude. 

These ideas were extended from cubic to quartic interactions in \cite{Banerjee:2018tun} where the underlying positive geometry was Stokes polytope. However unlike Associahedron, which is unique (in a given dimension), there are several Stokes polytopes in any given dimension and it was shown that one had to sum over canonical forms of all such polytopes to obtain scattering amplitude of $\phi^{4}$ theory. Not all Stokes polytopes contributed equally but one had to assign different weights to each Stokes polytope.  In \cite{Banerjee:2018tun} it was argued that these weights were not assigned to a given Stokes polytope but to an equivalence class of such polytopes which were related to each other by cyclic permutations and that in each such class, one could choose a representative that we called primitive. Whence the computation of scattering amplitude reduced to the problem of finding all the primitives and assigning weights to them. 

In this paper, continuing along the lines of \cite{Arkani-Hamed:2017mur,Banerjee:2018tun} we extend the Amplituhedron program to (tree-level) planar amplitudes for  massless scalar field theories with $\phi^{p}$ interactions. We have shown that the positive geometry underlying scattering amplitudes in this theory is a class of polytopes called accordiohedron. Accordiahedron is a family of polytopes whose members include associahedron and Stokes Polytope.\footnote{Our work thus shows that the positive geometry underlying planar amplitudes in any scalar field theory is an Accordioheron}

Just as in the case of quartic interactions there exists no single accordiohedron of a given dimension $n$ and a weighted sum of canonical forms of all the accordiohedra of a given dimension $n$ does indeed produce the full planar amplitude. This re-affirms and generalises the result we had obtained in in the case of quartic interactions. 

In \cite{Banerjee:2018tun} the problem of counting all primitives (corresponding to Stokes polytopes) of a given dimension had remained open. In this paper we fill this gap  and in fact extend this notion to primitive accordioheron. We give an enumeration of the number of primitives at arbitrary dimension $n$ and a complete classification of primitive diagrams for $n \le 3$. 

We then gave a prescription to compute the weights and provided the results for the weights obtained by using our prescription for all $p\ >\ 4$ in $n=1,2$ dimensions and $p \le 12 $ for $n=3$. 

Accordioheron is a very general polytope and one may wonder if they can be used to extend the Amplituhedron program to Scalar theories with mixed-vertices (e.g.  theories with cubic as well as quartic interactions). It turns out that this is indeed the case. \cite{mrunmay}

There are several outstanding questions that arise out of our analysis. In \cite{Salvatori:2018fjp} it was shown that the 1-loop integrand of $\phi^{3}$ theory also corresponds to canonical form on a polytope which is well known in mathematical literature called Halohedron. Whether this idea can be extended to 1-loop integrand of $\phi^{p}$ theories remains to be seen. 

One of the most striking results obtained in \cite{Arkani-Hamed:2017mur} was the derivation of CHY formula for bi-adjoint scalar $\phi^{3}$ interactions from the canonical form on kinematic space associahedron. More in detail, it was shown that the (tree-level) moduli space of CHY scattering equations admits a compactification which is nothing but an Associahedron (called worldsheet Associahedron)  and the Scattering equations can be understood as diffeomorphisms between the world-sheet Associahedron and the Associahedron in Kinematic space. It was also shown that the canonical form on Kinematic space Associahedron is nothing but a push-forward of the so-called Park Taylor form which is the canonical form on world-sheet associahedron. Although CHY integrands exist for $\phi^{p}$ interactions for $p\ >\ 3$ they do not admit any such geometric interactions. Our hope is that understanding of kinematic space Accordiohedron is the first step in ``geometrizing" the CHY formula for $\phi^{p}$ theories. 

\section*{Acknowledgements} 
I would like to thank Alok Laddha for many illuminating discussions, constant guidance and various valuable comments on improving the manuscript. I would also like to thank Nemani Suyanaryana, Sujay Ashok, Nima-Arkani Hamed, Song He and Guilio Salvatori and Pinaki Banerjee for insightful discussions and constant encouragement. I would also like to thank the participants of Soft Holography conference in pune and the string group at university of Torino where a preliminary version of this work was presented.

\appendix
\section{Counting planar diagrams}\label{appendixA}
In this appendix we want to derive an explicit formula for the number $a_n$ of planar tree level Feynman diagrams with $n$ vertices and arbitrary interactions. We begin by considering the ``toy" equations of motion which are equations of motion with a source where we set the d'Alembertian and couplings to one \cite{Cheung:2017pzi}. In other words this is equivalent to setting all propagators and vertices to one. Since, the perturbative solutions to equations of motion with a source are in 1-1 correspondence with connected tree amplitudes, the solutions of these equations gives us the generating functional for the number of Feynman diagrams.
\vskip .2cm
\noindent{\it Planar diagrams in $\phi^{p}$ theory}: The toy equations for this case are 
\bea \label{toy}
\phi = J + \phi^{p-1}
\eea 
We want to $\phi(J)$ from above which would be the generating function of Planar diagrams.
Note that we have $\phi^{p-1}$ instead of $\frac{\phi^{p-1}}{(p-1)!} $ since we only want to count planar diagrams which correspond to diagrams that have a fixed ordering.
We can invert (\ref{toy}) by treating it as a formal power series and using the Lagrange series inversion formula.

If $z=f(w)$ which is analytic about $w=0$ and $f'(0) \neq 0$ then $w=g(z)$ in the neighbourhood of $z=0$ with $g(z)$ given by the power series:
\bea
g(z) = \sum _{n=1}^{\infty} g_n \frac{(z-f(0))^n}{n!} \nonumber
\eea
where,
\bea
g_n =\frac{d^{n-1}}{d w^{n-1}} \left(\frac{w}{f(w)-f(0)}  \right)^n \bigg|_{w=0} \nonumber
\eea
Using this for (\ref{toy}) with $f(z) =z- z^{p-1}$ we get
\bea \label{genfn}
\phi(J) = \sum_{n=0}^{\infty} \left( (p-1) n \atop{n} \right) \frac{J^{(p-2)n+1}}{(p-2)n+1}  
\eea
The number of planar trees $a_n$ with $n$ vertices is the coefficient of $J^{n-1}$ in the series above which is the Fuss-Catalan number $F_{n,p}$.
\section{Proof of identity for Fuss-Catalan numbers}\label{B}
In this appendix we want to prove the identity (\ref{combiident}) which we had used in the sections on counting primitives. We begin by noting that if:
\bea \label{2iden}
 \tilde{F}_{n,p,r_1+r_2} &=& \sum_{k=0}^{n} \tilde{ F}_{k,p,r_1}  \tilde{F}_{n-k,p,r_2}  
 \eea
 Then we could recursively prove the general identity for example,
 \bea
  \sum_{k_1,k_2=0 \atop{k_1+k_2=n}}^{n} F_{k_1,p,r_1}  F_{k_2,p, r_2} F_{n-k_1-k_2, p ,r_3}  &=& \sum_{k=0}^{n} \sum_{k_1=0}^{k} \tilde{F}_{k_1,p,r_1}  \tilde{F}_{k-k_1,p,r_2} \tilde{F}_{n-k,p,r_3} \nonumber  \\  &=& \sum_{k=0}^{n}  \tilde{F}_{k,p,r_1+r_2} F_{n-k,p,r_3} \nonumber \\ &=& \tilde{F}_{n,p,r_1+r_2+r_3} \nonumber
 \eea
 Thus, we shall only need to show (\ref{2iden}), and substitute $r_1,r_2,...,r_i=1$ to get  (\ref{combiident}). We begin with the right hand side of  (\ref{2iden})
 {\small \bea \label{rhs}
 \sum_{k=0}^{n} \frac{r_1}{(p-2) +r_1} \left(  (p-1)k+r_1-1\atop{k}\right) \frac{r_2}{(p-2) (n-k)+r_2 } \left(  (p-1)(n-k)+r_2-1\atop{n-k}\right) 
 \eea }
 To evaluate this expression we shall use,
 \bea \label{trick}
 \left( n \atop{k} \right) = \frac{1}{2 \pi \bf{i}} \oint_{|z|=\epsilon} \frac{(1+z)^n}{z^{k+1}} dz 
 \eea 
 We can rearrange (\ref{rhs}) as
{\scriptsize \bea \label{rhs1}
&=& \frac{r_1}{(p-2)n+r_1+r_2} \sum_{k=0}^{n} \left((p-1) k+r_1 -1 \atop{k} \right) \left( \left((p-1) (n-k)+r_2 -1 \atop{n-k} \right)- (p-2)  \left((p-1) (n-k)+r_2 -1 \atop{n-k-1} \right)  \right) \nonumber\\ &+& \frac{r_2}{(p-2)n+r_1+r_2} \sum_{k=0}^{n}  \left( \left((p-1) k+r_2 -1 \atop{k} \right)- (p-2)  \left((p-1) k+r_2 -1 \atop{k-1} \right)  \right) \left((p-1) (n-k)+r_1 -1 \atop{n-k} \right) 
 \eea }
 We can now apply (\ref{trick}) to  each term in the above sum
 {\scriptsize \bea
 \sum_{k=0}^{n} \left((p-1) k+r_1 -1 \atop{k} \right) \left((p-1) (n-k)+r_2 -1 \atop{n-k} \right) &=&\frac{1}{(2\pi {\bf i})^2} \oint_{|w| =\gamma} \oint_{|z|=\epsilon}  \frac{(1+z)^{r_1 -1} (1+w)^{(p-1)n+r_2-1}}{z w^{n+1}} \nonumber \sum_{k=0}^{\infty} \left( \frac{w (1+z)^{p-1}}{z (1+w)^{p-1}} \right)^{k} dz dw\nonumber \eea}
  {\scriptsize \bea
 \sum_{k=0}^{n} \left((p-1) k+r_1 -1 \atop{k-1} \right) \left((p-1) (n-k)+r_2 -1 \atop{n-k} \right) &=&\frac{1}{(2\pi {\bf i})^2} \oint_{|w| =\gamma} \oint_{|z|=\epsilon}  \frac{(1+z)^{r_1 -1} (1+w)^{(p-1)n+r_2-1}}{ w^{n+1}} \nonumber \sum_{k=0}^{\infty} \left( \frac{w (1+z)^{p-1}}{z (1+w)^{p-1}} \right)^{k} dz dw \nonumber \eea}
 We have replaced the finite $k$ sum  by an infinite sum since 
{\scriptsize  \bea
 \left( n \atop{n-k} \right) = \frac{1}{2 \pi \bf{i}} \oint_{|z|=\epsilon} \frac{(1+z)^n}{z^{n-k+1}} dz \nonumber
 \eea }
 vanishes for $k>n$.  
 {\scriptsize \bea
 \sum_{k=0}^{n} \left((p-1) k+r_1 -1 \atop{k} \right) \left((p-1) (n-k)+r_2 -1 \atop{n-k} \right) &=&\frac{1}{(2\pi {\bf i})^2} \oint_{|w| =\gamma} \oint_{|z|=\epsilon}  \frac{(1+z)^{r_1 -1} (1+w)^{(p-1)n+r_2-1}}{w^{n+1}}  \frac{1}{\left(z-\frac{w(1+z)^{p-1}}{(1+w)^{p-1}}\right)} dz dw \nonumber \eea}
 The only pole in $z$ is at $z=w$, assuming $\epsilon > \gamma$ and thus by performing the z integral we get:
  {\scriptsize \bea
 \sum_{k=0}^{n} \left((p-1) k+r_1 -1 \atop{k} \right) \left((p-1) (n-k)+r_2 -1 \atop{n-k} \right) &=&\frac{1}{2\pi {\bf i}} \oint_{|w| =\gamma}  \frac{ (1+w)^{(p-1)n+r_1+r_2-1}}{w^{n+1}(1-(p-2)w)} dw \nonumber \eea}
 where, we used the fact that the residue of $\frac{(1+z)^{r_1-1}}{\left(z- \frac{w(1+z)^{p-1}}{(1+w)^{p-1}} \right)}$ at $z=w$ is given by $\tfrac{(1+w)^r_{1}}{1-(p-2)w}$\\
 Similarly, we have 
 {\scriptsize \bea
 \sum_{k=0}^{n} \left((p-1) k+r_1 -1 \atop{k-1} \right) \left((p-1) (n-k)+r_2 -1 \atop{n-k} \right) &=&\frac{1}{2\pi {\bf i}} \oint_{|w| =\gamma}  \frac{ (1+w)^{(p-1)n+r_1+r_2-1}}{w^{n}(1-(p-2)w)} dw \nonumber \eea}
 Substituting into (\ref{rhs1}) we get
 {\scriptsize \bea 
\sum_{k=0}^{n} \tilde{ F}_{k,p,r_1}  \tilde{F}_{n-k,p,r_2} &=& \frac{r_1+r_2}{(p-2)n+r_1+r_2} \oint_{|w|=\gamma}  \frac{(1+w)^{(p-1)n+r_1+r_2-1}}{w^{n+1}} dw \nonumber\\ &=&  \frac{r_1+r_2}{(p-2)n+r_1+r_2}\left( (p-1)n+r_1+r_2-1 \atop{n}\right)  \nonumber \\ &=&\tilde{F}_{n,p,r_1+r_2}\eea}
 
\bibliographystyle{utphys}
\bibliography{final}
\end{document}